\documentclass[english,aps,prl,preprint]{revtex4-1}
\usepackage[T1]{fontenc}
\usepackage[latin9]{inputenc}
\usepackage{amsmath}
\usepackage{amssymb}
\usepackage{graphicx}
\usepackage{color}
\makeatletter
\@ifundefined{textcolor}{}
{%
 \definecolor{BLACK}{gray}{0}
 \definecolor{WHITE}{gray}{1}
 \definecolor{RED}{rgb}{1,0,0}
 \definecolor{GREEN}{rgb}{0,1,0}
 \definecolor{BLUE}{rgb}{0,0,1}
 \definecolor{CYAN}{cmyk}{1,0,0,0}
 \definecolor{MAGENTA}{cmyk}{0,1,0,0}
 \definecolor{YELLOW}{cmyk}{0,0,1,0}
}

\usepackage{babel}
\newcommand{\ei}{\mbox{Ei}}

\makeatother

\usepackage{babel}
\begin{document}

\title{Keldysh field theory of a driven dissipative
Mott insulator: nonequilibrium response and phase transitions}

\author{S. Sankar and V. Tripathi}

\affiliation{Department of Theoretical Physics, Tata Institute of Fundamental
Research, Homi Bhabha Road, Navy Nagar, Mumbai 400005, India}

\date{\today}
\begin{abstract}
Understanding strongly correlated systems driven out of equilibrium is a challenging 
task necessitating the simultaneous treatment of quantum mechanics,
dynamical constraints and strong interactions. A Mott insulator subjected to a uniform and 
static electric field is prototypical, raising key questions such as the fate of Bloch oscillations 
with increasing correlation strength, the approach to a steady state DC transport regime and the role 
of dissipation in it, and electric field driven phase transitions. Despite tremendous efforts over the
last decade employing various numerical and analytical approaches, the manner in which
a nonequilibrium steady state gets established has remained an unresolved problem. 
We develop here an effective large-$\mathcal{N}$ Keldysh field theory
for studying nonequilibrium transport in a regular one-dimensional
dissipative Mott insulator system subjected to a uniform electric
field. Upon abruptly turning on the electric field (a quench), a transient
oscillatory current response reminiscent of Bloch oscillations is
found. In the regime of small tunneling conductance the amplitude of these oscillations, over a large time window, decreases as an inverse square power-law in time, ultimately going over to an exponential decay beyond a 
large characteristic time $\tau_{d}$ that increases with $\mathcal{N}.$ Such a relaxation to a steady state DC response is 
absent in the dissipation free Hubbard chain at half filling. The steady state current
at small fields is governed by large distance cotunneling, a process
absent in the equilibrium counterpart. The low-field DC current has
a Landau-Zener-Schwinger form but qualitatively differs from the expression
for pair-production probability for the dissipation free counterpart.
The breakdown of perturbation theory in the Mott phase  possibly signals a nonequilibrium
phase transition to a metallic phase.
Our study sheds light on the approach of a driven, dissipative strongly 
correlated system to a nonequilibrium steady state and also provides a
general analytic microscopic framework for understanding other nonequilibrium phenomena
in these systems.

\end{abstract}
\maketitle

\section{Introduction\label{sec:Introduction}}

A central challenge in the area of dissipative quantum systems driven far from equilibrium relates to understanding the 
relaxation of initial conditions and the approach to nonequilibrium steady states.
The temporal evolution is governed by the distribution of the initial disturbance over the many-body eigenmodes of the system, 
the nature of the bath and its coupling to the system, and the driving protocol.
Mott insulator systems driven out of equilibrium are particularly interesting as they provide a meeting ground
for quantum mechanics, strong interactions, dynamical processes and
constraints. Many recent studies have attacked the problem of the nonequilibrium response of fermionic 
\cite{fukui1998breakdown,oka2003breakdown,oka2005,oka2010dielectric,
eckstein2010dielectric,aron2012dielectric,tripathi2016parity,eckstein2011damping,okamoto2007nonequilibrium,freericks2006nonequilibrium,li2015electric,
freericks2008quenching,dias2007frequency,amaricci2012approach,aron2012dimensional,rozenberg2010mechanism,murakami2018nonequilibrium,
dutta2007effect,kirino2010nonequilibrium,heidrich2010nonequilibrium,oka2012nonlinear,aoki2014nonequilibrium,eckstein2013dielectric}
or bosonic \cite{sachdev2002mott,buchleitner2003interaction,carrasquilla2013scaling,trotzky2012probing,lankhorst2018scaling}
Mott insulator systems subjected to a uniform and static electric field. One of the key questions concerns the fate of Bloch oscillations 
with increasing correlation strength\,\cite{eckstein2010dielectric,eckstein2011damping,sachdev2002mott,carrasquilla2013scaling,freericks2006nonequilibrium,
freericks2008quenching,dias2007frequency,aoki2014nonequilibrium}.  
Another important question is regarding the role played by dissipation in the attenuation of the Bloch oscillations and the eventual
approach to a nonequilibrium steady state (DC transport in particular) \cite{li2015electric,amaricci2012approach,aron2012dielectric,tripathi2016parity,
aoki2014nonequilibrium,aron2012dimensional}. 
A third crucial issue is related to the nature of nonequilibrium phase transitions
in Mott insulator systems \cite{sachdev2002mott,oka2003breakdown,oka2010dielectric,oka2005,aron2012dielectric,tripathi2016parity,rozenberg2010mechanism,
oka2012nonlinear,eckstein2013dielectric,kirino2010nonequilibrium,heidrich2010nonequilibrium,oka2012nonlinear,aoki2014nonequilibrium}. 
Different techniques have been
employed in the literature that address some of these issues -- these include numerical approaches such as solving time-dependent Schr\"{o}dinger 
equations \cite{oka2003breakdown}, nonequilibrium dynamical mean-field theory (NDMFT) \cite{schmidt2002nonequilibrium,okamoto2007nonequilibrium,
li2015electric,freericks2008quenching,dias2007frequency,amaricci2012approach,aron2012dielectric,aron2012dimensional,
eckstein2013dielectric,aoki2014nonequilibrium,eckstein2010dielectric,eckstein2011damping},
 time dependent density matrix renormalization group (TDMRG) \cite{oka2005,dutta2007effect,kirino2010nonequilibrium,heidrich2010nonequilibrium}, 
 as well as analytic ones based on the Bethe ansatz \cite{oka2010dielectric,oka2012nonlinear}, including the
phenomenological generalizations to $\mathcal{PT}$-symmetric models \cite{tripathi2016parity,fukui1998breakdown}.
In this paper, we develop a new analytic field theoretical approach based on the Keldysh technique 
and address the above three questions. Our method also provides a general analytic 
framework to investigate novel and wide variety of nonequilibrium phenomena in strongly correlated systems.

It is long known that a noninteracting particle hopping on a periodic
lattice subjected to a uniform electric field exhibits Bloch oscillations
- the spectrum is discrete (Wannier-Stark ladder \cite{wannier1960wave,fukuyama1973tightly}),
and the particle motion is bounded. 
Correlations,
dissipation and disorder can all suppress the Bloch oscillations by
providing relaxation or breaking lattice translation symmetry. For
field strengths such that the potential energy change between neighboring
sites far exceeds correlation and other energy scales in the problem,
Bloch oscillations have been found to persist \cite{buchleitner2003interaction,carrasquilla2013scaling,eckstein2010dielectric,eckstein2011damping}.
Physically, this can be understood from the fact that the noninteracting
Wannier-Stark states are highly localized at the lattice sites at
strong fields, and the correlations remain local in the Wanner-Stark
basis. At fields where the potential energy drop in a bond is comparable
to the interaction strength, study of the Bose-Hubbard model at integer
filling establishes that the motion remains finite \cite{sachdev2002mott}.
Recent numerical studies of fermionic Mott insulators show that at large fields, the electrons execute 
Bloch oscillations whose frequency approaches the noninteracting counterpart \cite{eckstein2010dielectric}. At smaller fields, 
the understanding for a long time was that interactions, through mixing of different momentum modes, attenuate the Bloch oscillations ultimately giving way
to a steady state DC response\cite{eckstein2010dielectric,eckstein2011damping}. However recent work suggests that the 
 apparent steady state DC behavior is only transient and ultimately gives way to finite (oscillatory) motion
with a period different from that of the noninteracting Wannier-Stark
states \cite{aoki2014nonequilibrium}. The current understanding is that dissipation is a necessary ingredient for establishing steady state DC response.

Bloch oscillations can be suppressed by dissipation through coupling the system to a
bath. Earlier literature shows that even at a single-particle level,
coupling the system to a phonon bath \cite{emin1987phonon} or a fermionic bath \cite{han2013solution}
results in a finite DC response at any value of the coupling strength; however for the case of coupling to a phonon bath,
signatures of the Wannier-Stark ladder are still evident
in the spectral function, which are found to diminish with increasing
electron-phonon coupling \cite{cheung2013phonon}.
Recent works have also considered the effect of correlations in dissipative models.
The dissipation is introduced either by coupling the system to a bath 
\cite{aron2012dielectric,aron2012dimensional,li2015electric,amaricci2012approach} or
by phenomenological means, for example, by introducing non-Hermitian terms
in Hamiltonians preserving $\mathcal{PT}$ symmetry \cite{tripathi2016parity,fukui1998breakdown} or using Lindblad
formulations \cite{arrigoni2013nonequilibrium}. The former (heat bath) case has been studied using a
numerical Keldysh DMFT approach \cite{aron2012dielectric,aron2012dielectric}, while
the Bethe ansatz method is usually employed in the latter for one-dimensional
systems \cite{tripathi2016parity}. Both these approaches yield a
steady state nonequilibrium response and nonequilibrium transitions
from the Mott insulator state to a metallic state. In addition, an
important observation was made in Ref. \cite{aron2012dielectric}
that weak dissipation does not completely suppress quantum coherent
oscillations - the numerically calculated single particle spectral
function shows ``Bloch islands'' at beating frequencies involving
the noninteracting Bloch oscillations and the Coulomb interaction
strength. These features get suppressed as dissipation is increased.
Despite these advances in the numerical studies of the microscopic
model, many important issues have not yet been addressed; for instance, it
is not known how the transient Bloch oscillations decay in time eventually establishing a DC current state, and how they get
suppressed in the presence of dissipation. Phenomenological models such as the $\mathcal{PT}$
symmetric Hubbard models are analytically tractable and give valuable
insights such as the critical behavior near the nonequilibrium Mott
insulator to metal transition; however relating the model parameters
directly to experimentally relevant quantities has proved to be a
challenge. Moreover, these models are designed to study the nonequilibrium
steady state but not the transient response.

In band insulators, the linear response conductivity vanishes at zero
temperature but electronic transport at finite electric fields is
possible through the generation of low-energy particle-hole pairs
by the Landau-Zener-Schwinger (LZS) mechanism \cite{landau1932theorie,zener1934theory,schwinger1951gauge}, with the
probability $P$ of this process related to the electric field measured
in terms of the potential energy drop, $D,$ across a link, and the
band-gap $\Delta$ as $P\sim\exp[-\Delta^{2}/cD],$ where $c$ is
a constant with the dimension of energy. For the fermionic Hubbard
chain subjected to an electric field, a similar expression has been
proposed in Ref. \cite{oka2005}, with band-gap $\Delta$ being replaced
by the Mott gap. Turning on a finite dissipation (coupling to a fermionic
bath) under such nonequilibrium conditions, DMFT calculations of Ref.
\cite{aron2012dielectric} show that the Hubbard bands leak into the
Mott gap, and beyond some value of the dissipation strength, a quasiparticle
feature, signaling a bad metallic phase appears, in the spectral function.
The crucial question here is whether and under what circumstances
this dielectric breakdown becomes a true nonequilibrium phase transition.
Analysis of the phenomenological $\mathcal{PT}$ symmetric fermionic
Hubbard chain \cite{tripathi2016parity} suggests that this is a true
nonequilibrium quantum phase transition and is associated with breaking
of $\mathcal{PT}$ symmetry in the metallic phase. 

In this paper we develop an effective Keldysh field theory of a dissipative
one-dimensional Mott insulator subjected to a uniform electric field
and study it analytically to address the broad questions outlined
above. Our microscopic model consists of a one-dimensional array of
mesoscopic metallic quantum dots - each of these quantum
dots contains a large number of electrons occupying the dot energy
levels. The large number of degrees of freedom (DoF) in each mesoscopic
dot effectively constitute a fermionic bath and provide a source of
dissipation through the Landau damping mechanism. In addition, as
we discuss below, the large DoF acts as a large-${\cal N}$ parameter
(see also \cite{zarand2000two}) and facilitates a tractable analytic
treatment of our model. The analytic tractability that our large-${\cal N}$
formulation provides is analogous to that of large dimensionality
in the DMFT approach to the Hubbard model. Under equilibrium conditions,
the model is described by the following Hubbard-like Hamiltonian with
multiple flavors (representing dot energy levels) of electrons at
each site (we set electron charge $e$ = 1, lattice spacing $a$ =
1, $\hbar=1$, $k_{B}=1$): 
\begin{eqnarray}
\hat{H} & = & \hat{H}^{0}+\hat{H}^{C}+\hat{H}^{T},\,\text{where}\label{eq:fermi_model}\\
\hat{H}^{0} & = & \sum_{k,\alpha}\xi_{\alpha}c_{j,\alpha}^{\dagger}c_{j,\alpha},\\
\hat{H}^{C} & = & \sum_{k}E_{C}\left[\left(\sum_{\alpha}c_{k,\alpha}^{\dagger}c_{k,\alpha}\right)-N_{0}\right]^{2},\\
\hat{H}^{T} & = & \sum_{k}\sum_{\alpha,\beta}\left(\tilde{t}_{\alpha\beta}^{k,k+1}c_{k,\alpha}^{\dagger}c_{k+1,\beta}+\text{h.c.}\right).
\end{eqnarray}
Here $k$ labels the site index, $\alpha$ represents the different
energy levels ($E_{\alpha}$) within a dot, $\xi_{\alpha}=E_{\alpha}-\mu$
($\mu$ being the Fermi level in the dot), $\tilde{t}_{\alpha\beta}^{k,k+1}$
is the inter-dot tunneling matrix element connecting levels $\alpha$
and $\beta$ on dots labeled $k$ and $k+1$ respectively, $E_{C}$
is the Coulomb energy of single-electron charging, and $N_{0}$ is
the equilibrium charge on a dot. The tunneling between the dots could
be through an insulating barrier (as is the case in granular metals)
or through ballistic point contacts (as may be the case in artificial
quantum dot arrays). The Fermi energy in each dot is assumed to be
the largest energy scale. In addition, we also have a small energy
scale, $\delta,$ which is the mean level spacing in the dot and is
approximately related to the volume of the dot, $V,$ and the density
of states at the Fermi level, $\nu(\mu)$ through $\delta\approx1/(\nu(\mu)V).$
Elementary excitations in each isolated dot are of the low-energy
particle-hole kind, which in the limit of large dot size, tend to
become gapless. Interestingly, other models such as the Sachdev-Ye-Kitaev
(SYK) \cite{sachdev1993gapless,kitaev2015simple} model on a one-dimensional
lattice \cite{gu2017local,davison2017thermoelectric} interaction
share a similar structure, and are also characterized by gapless excitations
locally. 

We study the model in Eq. (\ref{eq:fermi_model}) in the Mott insulator
regime where $E_{C}\gg\delta,\,T$ and $g\lesssim1,$ where $T$ is
the temperature, and $g$ is the dimensionless inter-dot tunneling
conductance. For granular metals, the intergrain tunneling conductance
is of the form $g\approx\pi^{2}|\tilde{t}_{\alpha,\beta}|^{2}(V\nu(\mu))^{2}=\pi^{2}|\tilde{t}_{\alpha,\beta}|^{2}/\delta^{2}.$
For ballistic point contacts separating the quantum dots, the transverse
(waveguide) momentum $\mathbf{k}_{\perp}$ is conserved during tunneling
(i.e. $\tilde{t}_{\alpha,\beta}\equiv\tilde{t}_{\mathbf{k}_{\perp}}$)
but the longitudinal momentum $k_{\parallel}$ is not, and $g$ has
the form, $g\approx\pi^{2}\sum_{\mathbf{k}_{\perp}}|\tilde{t}_{\mathbf{k}_{\perp}}|^{2}(\nu^{1D}L)^{2},$
where $\nu^{1D}$ is the one-dimensional density of states associated
with the different sub-bands labeled by $\mathbf{k}_{\perp}$ and
$L$ is the dot size. In this Mott insulator regime, a conventional
perturbation expansion in the interaction is not possible. We therefore
adopt a bosonization scheme well known in the literature as the Ambegaokar-Eckern-Schön
(AES) \cite{ambegaokar1982quantum,beloborodov2007granular} model
of granular metals - a class of Mott insulators. The AES model is,
in effect, a rotor model with the difference that now the phases at
each site in the AES model are dual to the total charge in the dot
at that site. The AES model consists of a charging part that represents
Coulomb blockade effects, and a dissipative tunneling part that describes
inter-dot hopping of electrons. Unlike other dissipative models such
as Caldeira-Leggett \cite{caldeira1981influence}, the tunneling part
of the AES model is periodic in the phase fields reflecting charge
quantization. The large number of degrees of freedom on each dot makes
the model analytically tractable, allowing one to discard terms in
the effective action that are higher order than two in the inter-dot
tunneling conductance. The model is tailor-made for studying transport,
and consequently, information about the internal low-energy excitations
at a site appears only at the level of the tunneling term.

In equilibrium or linear response situations, the AES model appears
in diverse contexts including unusual transport phenomena in granular
Mott insulators such as cotunneling dominated variable-range hopping
\cite{tran2005multiple,beloborodov2007granular} and breakdown of
the Wiedemann-Franz law by emergent bosonic modes \cite{tripathi2006thermal}
and the Kondo effect in quantum critical metals \cite{sengupta2000spin,loh2005magnetic}.
A bosonic channel for thermal transport analogous to that in the AES
model \cite{tripathi2006thermal} has recently been reported for the
SYK model \cite{davison2017thermoelectric}. It is also well-known
that even in the regime of metal-like conduction ($g\gg1,$ $T\gg g\delta$),
the low-energy excitations of the AES model are not quasiparticle-like,
i.e., are not characterized by their momenta and spin, a property
shared with the SYK model \cite{davison2017thermoelectric}.

We generalize the AES model to the nonequilibrium case using the Keldysh
formalism. For the case of a single mesoscopic quantum dot connected
to noninteracting leads, a similar Keldysh generalization has been
studied in the literature (see e.g. \cite{altland2010condensed}).
The granular chain, as we shall see, has significantly different physics
from the single dot problem arising from the periodicity of the lattice
and also the relevance of long-range tunneling processes since potential
energy gain from cotunneling over multiple dots can offset the Coulomb
blockade effects. In the equilibrium (Matsubara) treatment of the
AES model, in order to properly treat charge quantization effects,
essential in Coulomb blockade, finite winding numbers of the phase
fields must be taken into account. In the real time Keldysh case,
this is achieved by going to a mixed phase-charge representation (instead
of a pure phase-only representation) and restricting the path integral
over the classical component of the charge field to integer values.

We calculate the current response of our Keldysh AES model for the
granular Mott insulator subjected to a uniform electric field at temperatures
much smaller than $D$ and $E_{C},$ and we further assume the mesoscopic
dots are sufficiently large so that the temperature greatly exceeds
the mean level spacing $\delta.$ After the electric field is switched
on, the leading order (in $g$) current response shows an oscillatory
transient response whose primary components are the two beat frequencies,
$\omega_{\pm}=|D\pm2E_{c}|,$ 
\begin{align}
J_{\text{tr}} & \approx 
- \frac{4g\Theta(\tau)}{(2\pi)^{2}E_{C}}
\frac{1}{\tau^{2}} 
\left[
\left(\frac{D-2E_{C}}{D+2E_{C}}\right)
\sin((D+2E_{C})\tau) +
\left(\frac{D+2E_{C}}{D-2E_{C}}\right)
\sin((D-2E_{C})\tau)
                 \right].
\label{eq:Jtrans-final_intro}
\end{align}
These oscillations arise, as we shall
show in the paper, from a combination of the periodicity of the lattice,
Coulomb correlations, and charge quantization. These beat frequencies
have also been observed \cite{aron2012dielectric} in DMFT calculations
of the dissipative Hubbard model in the form of ``island'' features
in the spectral function, and in the dissipationless Bose-Hubbard
model \cite{buchleitner2003interaction}. In the absence of correlations
($E_{C}=0$), these oscillations would correspond to the Bloch oscillation
frequency $\omega_{B}=|D|.$
However, the $1/\tau^{2}$ decay of the amplitude of the current oscillations does not persisit indefinitely, and we show that it crosses over to an exponential decay to the steady state beyond a characteristic time $\tau_{d} \sim 1/T_D,$ where $T_D$ is the effective electron temperature in the dots in the nonequilibrium steady state. We find that the temperature $T_D$ decreases with $\mathcal{N},$ and vanishes as $\mathcal{N}\rightarrow\infty.$ 
 
 Apart from these oscillations, the current also has a finite DC component
for $|D|>2E_{C},$
\begin{align}
J_{\text{dc}} & =\frac{g\Theta(\tau)}{\pi}\left[(D-2E_{C})\Theta(D-2E_{C})+(D+2E_{C})\Theta(-2E_{C}-D)\right], \label{eq:dc-response_intro}
\end{align} 
and is a direct consequence of the presence of
dissipation.

Next, to understand the nature of the DC response at small fields,
$|D|<2E_{C},$ we consider the long time limit of the current response.
For this purpose, we take into account higher order cotunneling processes
over multiple dots such that the Coulomb blockade is offset by the
extra potential energy gain. We provide analytic expressions for the
field dependence of current up to $O(g^{2}).$ The analysis of higher
order terms at arbitrary field strengths rapidly becomes very complicated;
however we infer some general features. In the zero temperature limit,
there is a hierarchy of thresholds, $D_{\text{th}}^{(n)}=2E_{C}/n,$
with the $n^{\text{th}}$ order current corresponding to the matching
of the Coulomb scale with the electrostatic potential energy gain
from cotunneling over $n$ successive dots. The leading order in $g$
contributions to the current near these thresholds has the form 
\begin{align}
j^{(n)}(D) & \sim nDg^{n}(1-2E_{C}/nD)^{2n-1}\Theta(nD-2E_{C}),\label{eq:nth_order_current}
\end{align}
where $\Theta$ is the Heaviside step function. Based on this expression,
we show that at low fields and small $g,$ the field dependence of
the current has the LZS form, $j(D)\sim D[g/\ln^{2}(1/g)]^{2E_{C}/D},$
but with qualitative differences from the LZS particle-hole pair production
probability $P\sim e^{-E_{C}^{2}/cD}$ for the non-dissipative Hubbard
chain at half filling \cite{oka2010dielectric} deep in the Mott insulator
phaes. 

An important question relates to the nature of the transition from
the Mott insulating state to a conducting state as a function of the
field. In the dissipation free case, it is evident from the expression
for the LZS pair production probability that it is a crossover, howsoever
sharp, and not a true phase transition. A true phase transition to
a metallic state is indicated if the perturbation expansion for the
current made from within the Mott insulator phase diverges as a function
of $g(\lesssim1)$ or $D(<2E_{C}).$ If the form of the current is
assumed to have the form shown in Eq. (\ref{eq:nth_order_current})
for a finite but small field strength away from the thresholds, then
the criterion for divergence of the perturbation expansion for the
current is 
\begin{align}
g\exp[D/E_{C}] & \sim1.\label{eq:pert-th-breakdown}
\end{align}
However, as we have already mentioned above, the field dependence
of high-$n^{\text{th}}$ order terms is complicated for fields away
from the respective thresholds $D^{(n)}=2E_{C}/n,$ and it is not
currently clear to us how the above criterion would change.

The rest of the paper is organized as follows. In Sec. \ref{sec:Keldysh-AES-action},
beginning with the microscopic model of Eq.~(\ref{eq:fermi_model}),
we outline the derivation of our effective Keldysh-AES action. The
electric field is introduced through a time-dependent vector potential.
We also present the functional representation of the charge current
in terms of the correlation functions of the phase fields. In Sec. \ref{sec:Transient-current-response},
we analyze the leading order contribution to the
current from the time the electric field is turned on. We show that
there are Bloch-like oscillations whose amplitudes decay as a power-law
in time upto a large time $\tau_{d}$. Further, the existence of a finite DC response at long times
is also established. Sec. \ref{sec:Steady-state} is devoted to the analysis
of the long-time DC behavior for small field strengths. For this purpose,
the higher order cotunneling processes over multiple dots are considered
in a perturbative expansion in small $g,$ around the ``atomic limit''
of isolated dots. We discuss the LZS form of the current response
at small fields, and the possible nonequilibrium phase transition
to a metallic state. Finally, in Sec. \ref{sec:discussion} we conclude
with a discussion of our results and open questions.

\section{Keldysh-AES action\label{sec:Keldysh-AES-action}}

In this Section, we obtain the effective Keldysh-AES action from the
microscopic Hamiltonian introduced in Eq. (\ref{eq:fermi_model})
and also provide functional representation of the charge current that
will be used throughout. Our derivation of the effective Keldysh-AES
action parallels the one in Ref. \cite{altland2010condensed} for
the case of a single quantum dot connected to noninteracting leads. 

The first step consists of Hubbard-Stratonovich decoupling of the
part of the action corresponding to Eq. (\ref{eq:fermi_model}) that
contains the Coulomb interaction term: 
\begin{eqnarray}
e^{-i\int_{t}H_{C}} & = & \exp\left[-\iota\sum_{k}\int_{t}E_{C}\left(\sum_{\alpha}\bar{\psi}_{k,\alpha}\psi_{k,\alpha}-N_{0}\right)\left(\sum_{\alpha}\bar{\psi}_{k,\alpha}\psi_{k,\alpha}-N_{0}\right)\right]\nonumber \\
 & \propto & \int DV\exp\left[\iota\sum_{k}\int_{t}\frac{1}{4E_{C}}\left(V-2E_{C}\left(\sum_{\alpha}\bar{\psi}_{k,\alpha}\psi_{k,\alpha}-N_{0}\right)\right)^{2}\right]e^{-i\int_{t}H_{C}}
\end{eqnarray}
To study nonequilibrium transport, we put our action on the Keldysh
contour and we label the fields with superscripts $+$ and $-$ corresponding
respectively to the forward and backward time parts of the Keldysh
contour. For incorporating the initial condition information (i.e.
the initial density matrix) it is customary to work with a rotated
classical-quantum basis in the Keldysh space: 
\begin{eqnarray}
V^{c} & = & \frac{1}{2}(V^{+}+V^{-})\mbox{ , }V_{q}=V^{+}-V^{-},\\
\psi^{c} & = & \frac{1}{\sqrt{2}}(\psi^{+}+\psi^{-})\mbox{ , }\psi^{q}=\frac{1}{\sqrt{2}}(\psi^{+}-\psi^{-}),\\
\bar{\psi}^{c} & = & \frac{1}{\sqrt{2}}(\bar{\psi}^{+}-\bar{\psi}^{-})\mbox{ , }\bar{\psi}^{q}=\frac{1}{\sqrt{2}}(\bar{\psi}^{+}+\bar{\psi}^{-}),\\
\Psi & = & \begin{pmatrix}\psi^{c}\\
\psi^{q}
\end{pmatrix}\mbox{ , }\bar{\Psi}=\begin{pmatrix}\bar{\psi}^{c} & \bar{\psi}^{q}\end{pmatrix}.
\end{eqnarray}
We call the superscripts $c$ and $q$ the ``classical'' and ``quantum''
components respectively. The action $S$ now assumes the form, 
\begin{align}
S & =S^{0}+S^{C}+S^{T},\,\text{where}\nonumber \\
S^{0} & =\sum_{k,\alpha}\int_{t}\bar{\Psi}_{k,\alpha}\begin{bmatrix}\iota\partial_{t}+\iota\eta+\mu-E_{\alpha}-V_{k}^{c} & -\frac{V_{k}^{q}}{2}+2\iota\eta F_{k}\\
-\frac{V_{k}^{q}}{2} & \iota\partial_{t}-\iota\eta+\mu-E_{\alpha}-V_{k}^{c}
\end{bmatrix}\Psi_{k,\alpha},\nonumber \\
S^{C} & =\sum_{k}\int_{t}\left(\frac{1}{2E_{c}}V_{k}^{c}V_{k}^{q}+N_{0}V_{k}^{q}\right),\nonumber \\
S^{T} & =\sum_{k,\alpha,\beta}\int_{t}\bar{\Psi}_{k\alpha}\begin{bmatrix}\tilde{t}_{\alpha,\beta}^{k,k+1} & 0\\
0 & \tilde{t}_{\alpha,\beta}^{k,k+1}
\end{bmatrix}\Psi_{k+1,\beta}+\text{c.c.}\label{eq:eff-action1}
\end{align}
Here $F_{k}$ is related to the distribution function for \emph{noninteracting
}electrons in the $k^{\text{th}}$ dot and is, in general, a function
of two time arguments, i.e., $F_{k}(t,t').$ For the case of thermal
equilibrium, $F_{k}$ depends only on the difference $t-t',$ and
in frequency space, it has the form $F(\omega)\equiv1-2f(\omega)=\tanh(\omega/2T),$
where $f(\omega)$ is the Fermi-Dirac distribution function and $T$
is the temperature. The infinitesimally small positive constant, $\eta,$
ensures the theory has the proper causal structure. At this stage,
it would seem natural to integrate out the noninteracting fermions,
and expand the resulting determinant to obtain an effective field
theory for the Hubbard-Stratonovich fields. However, the Hubbard-Stratonovich
fields effectively shift the entire band of electrons and, in fact,
the shifts are large ($\sim E_{C}$) whenever tunneling events occur.
We therefore perform a gauge transformation to eliminate the fluctuating
Hubbard Stratanovich fields that appear in $S^{0}$
\begin{eqnarray}
\Psi_{k,\alpha} & \rightarrow & e^{-\iota\hat{\phi_{k}}}\Psi_{k,\alpha}\mbox{ , }\bar{\Psi}_{k,\alpha}\rightarrow\bar{\Psi}_{k,\alpha}e^{\iota\hat{\phi_{k}}},\label{eq:gauge-transf}
\end{eqnarray}
where 
\begin{align}
\hat{\phi}_{k} & =\phi_{k}^{c}+\phi_{k}^{q}\frac{\sigma_{1}}{2},\label{eq:hat-fields}
\end{align}
and the phase fields $\hat{\phi}_{k}$ are chosen such that their
classical and quantum components obey 
\begin{align}
\partial_{t}\phi_{k}^{c,q} & =V_{k}^{c,q}.\label{eq:gauge-choice}
\end{align}
After the above gauge transformation, we have, 
\begin{eqnarray}
S^{0} & = & \sum_{k,\alpha}\int_{t}\bar{\Psi}_{k,\alpha}\begin{bmatrix}\iota\partial_{t}+\iota\eta+\mu-E_{\alpha} & 2\iota\eta F_{k}\\
0 & \iota\partial_{t}-\iota\eta+\mu-E_{\alpha}
\end{bmatrix}\Psi_{k,\alpha}.\\
S^{C} & = & \sum_{k}\int_{t}\left(\frac{1}{2E_{c}}\partial_{t}\phi_{k}^{c}\partial_{t}\phi_{k}^{q}+N_{0}\partial_{t}\phi_{k}^{q}\right),\label{eq:SC-N0}\\
S^{T} & = & \sum_{k,\alpha,\beta}\int_{t}\left(\tilde{t}_{\alpha,\beta}^{k,k+1}\bar{\Psi}_{k\alpha}\exp(-\iota\hat{\phi}_{k,1})\Psi_{k+1,\beta}+\mbox{c.c.}\right)\mbox{ , }\hat{\phi}_{k,1}=\hat{\phi}_{k+1}-\hat{\phi}_{k},
\end{eqnarray}
The term in Eq. (\ref{eq:SC-N0}) proportional to $N_{0}$ is a Berry
phase term. Our next step is to integrate out the fermions to obtain
an effective action in terms of the phase fields. We denote the fermion-bilinear
part of the action as $S_{F}=S^{0}+S^{T}=\hat{\bar{\Psi}}\hat{G}^{-1}\hat{\Psi}$,
with 
\begin{equation}
\hat{G}^{-1}=\hat{G}_{0}^{-1}+\hat{T},\label{eq:g0+T}
\end{equation}
where 
\begin{eqnarray}
(\hat{G}_{0})_{k,\alpha;k,\alpha}^{-1} & = & \begin{bmatrix}(g_{k,\alpha}^{R})^{-1} & 2\iota\eta F_{k}\\
0 & (g_{k,\alpha}^{A})^{-1}
\end{bmatrix},\label{eq:G0-inv}\\
\hat{T}{}_{k,\alpha;k+1,\beta} & = & \tilde{t}_{\alpha,\beta}^{k,k+1}\exp(-\iota\hat{\phi}_{k,1}).
\end{eqnarray}
In Eq. (\ref{eq:G0-inv}), the diagonal elements are the usual inverse
retarded and advanced Green functions, 
\begin{align}
(g_{k,\alpha}^{R,A})^{-1} & =\iota\partial_{t}\pm\iota\delta+\epsilon_{F}-E_{\alpha}.\label{eq:g-RA}
\end{align}
The inter-dot hopping matrix $\hat{T}$ is diagonal in Keldysh space
as well as in the time indices. Integrating out the fermions gives
us $Z=\int D\phi\exp(\iota S^{C}[\phi]+\mbox{tr}\ln(\iota\hat{G}^{-1})),$
and we use Eq. (\ref{eq:g0+T}) to re-express the fermionic determinant
as 
\begin{equation}
\ln(\hat{G}^{-1})=\ln(1+\hat{G_{0}}\hat{T})+\ln(\hat{G}_{0}^{-1}).
\end{equation}
To obtain the effective action in terms of the phase fields, we discard
the $\phi$-independent $\ln(\hat{G}_{0}^{-1})$ make a Taylor expansion
of $\ln(1+\hat{G}_{0}\hat{T})$.\textbf{ }The first order term vanishes
since $\mbox{tr}(\hat{G}_{0}\hat{T})=0$ as $\hat{G}_{0}$ is diagonal
in $k$ and $T_{k;k}=0$. Then, up to second order in $\hat{T}$ we
have 
\begin{eqnarray}
Z & = & \int D\phi\exp(\iota S^{C}[\phi]+\iota S^{\mbox{tun}}[\phi])\mbox{ , }S^{\mbox{tun}}[\phi]=\frac{\iota}{2}\mbox{tr}\left(\hat{G}_{0}\hat{T}\hat{G}_{0}\hat{T}\right).\label{eq:-2nd-od}
\end{eqnarray}
Here $\hat{G}_{0}$ has the following structure in Keldysh space:
\begin{equation}
(\hat{G}_{0})_{k,\alpha;k,\alpha}(t,t^{'})=\begin{bmatrix}g_{k,\alpha}^{R} & F_{k}(g_{k,\alpha}^{R}-g_{k,\alpha}^{A})\\
0 & g_{k,\alpha}^{A}
\end{bmatrix}(t,t^{'}),
\end{equation}
where 
\begin{equation}
g_{k\alpha}^{R,A}(t,t^{'})=\frac{1}{2\pi}\int_{\omega}g_{k,\alpha}^{R,A}(\omega)\exp(-\iota\omega(t-t^{'}))=\int_{\omega}\frac{\exp(-\iota\omega(t-t^{'}))}{\omega\pm\iota\delta+\mu-E_{\alpha}}.
\end{equation}
We assume that the matrix elements of $\hat{T}$ are independent of
the energy indices and also replace summation over the discrete states
by corresponding integrals, $\sum_{\alpha}\leftrightarrow V\int_{\epsilon}d\epsilon\,\nu(\epsilon),$
with $\nu(\epsilon)=\frac{1}{V}\sum_{\alpha}\delta(\epsilon-E_{\alpha})$
the density of states in a dot. The summations over the energy indices
gives quantities of the form $\sum_{\alpha}g_{k,\alpha}^{R,A}(\omega)=V\int_{\epsilon}\nu(\epsilon)g_{k,\alpha}^{R,A}(\omega)\approx\mp(\pi\iota)V\nu(\omega+\mu)\approx\mp(\pi\iota)V\nu(\mu).$
With these approximations, we arrive at 
\begin{align}
\mbox{tr}(\hat{G}_{0}\hat{T}\hat{G}_{0}\hat{T}) & \approx-2\pi^{2}|\tilde{t}|^{2}(V\nu(\mu))^{2} \int_{t,t^{'}}\sum_{k}\mbox{tr}\left[\Lambda_{k}(t-t^{'})\exp(-\iota\hat{\phi}_{k,1}(t^{'})) \right. \nonumber \\
& \qquad \qquad \qquad \qquad \qquad \qquad \qquad \left.\Lambda_{k+1}(t^{'}-t)\exp(\iota\hat{\phi}_{k,1}(t))\right],
\end{align}
where 
\begin{equation}
\Lambda_{k}(\omega)=(2\iota)\begin{bmatrix}G^{R}(\omega) & F_{k}(\omega)[G^{R}-G^{A}]\\
0 & G^{A}(\omega)
\end{bmatrix}\mbox{ , }G^{R,A}(\omega)=\frac{1}{2\pi}\int_{\epsilon}g_{k,\epsilon}^{R,A}(\omega).
\end{equation}
Thus, 
\begin{equation}
S_{\mbox{tun}}\approx-\iota g\int_{t,t^{'}}\sum_{k}\mbox{tr}\left[\Lambda_{k}(t-t^{'})\exp(-\iota\hat{\phi}_{k,1}(t^{'}))\Lambda_{k+1}(t^{'}-t)\exp(\iota\hat{\phi}_{k,1}(t))\right].\label{eq:Stun-2}
\end{equation}
For a granular metal, we assume that the tunneling matrix connects
any pair of levels in the neighboring grains with characteristic magnitude
$|\tilde{t}|,$ in which case, $g=\pi^{2}(V\nu(\mu))^{2}|\tilde{t}|^{2}\sim|\tilde{t}|^{2}(\mathcal{N}/\mu)^{2}.$
Here $g$ is the dimensionless inter-dot tunneling conductance and
$\mathcal{N}$ the total number of electrons in a dot. To give an
estimate of the largeness of $\mathcal{N},$ for a $10{\rm nm}$ metallic
dot with conduction electron density of $\sim10^{28}{\rm m}^{-3},$
we have $\mathcal{N}\sim10^{4}.$ Our regime of interest is $g\lesssim1,$
independent of the number of electrons in the dot. Thus for the granular
metal we require the tunneling amplitudes to scale as $|\tilde{t}|\sim1/\mathcal{N}.$
Physically, this means that as the number of transmission channels
increases, the individual tunneling amplitudes should scale inversely
so as to keep $g$ unchanged. 

For the case of ballistic point contacts, we label the energy levels
by transverse and longitudinal momenta, $\mathbf{k}_{\perp}$ and
$k_{\parallel}$ respectively. The transverse momentum is conserved
during tunneling but the longitudinal momentum is not. The tunneling
matrix element thus connects any pair of longitudinal momenta, and
we assume they all have a characteristic magnitude $|\tilde{t}|.$
In this case, the dimensionless conductance $g=\pi^{2}\sum_{\mathbf{k}_{\perp}}|\tilde{t}|^{2}(\nu^{1D}L)^{2}\sim|\tilde{t}|^{2}N_{\text{ch}}(\mathcal{N}_{1D}/\mu)^{2},$
where $N_{\text{ch}}$ is the total number of transverse channels
and $\mathcal{N}_{1D}$ is the typical number of electrons having
the same transverse momentum. To keep $g\lesssim1,$ we require the
tunneling amplitude to scale as $|\tilde{t}|\sim1/(\sqrt{N_{\text{ch}}}\mathcal{N}_{1D}),$
and we show below that the large-$\mathcal{N}$ parameter in this
case is $\mathcal{N}=N_{\text{ch}}.$

We will present below a large-$\mathcal{N}$ justification for dropping
higher order terms in the tunneling action.

\subsection{Consequences of large-$\mathcal{N}$ }

Let us now discuss a couple of crucial consequences of having a large
number of electrons in each dot. Consider first the $O(\tilde{t}^{4})$
term in the tunneling action for the granular metal. The basic argument
for disregarding such contributions has been presented in Ref \cite{beloborodov2007granular}
. Here we show that this is essentially a large-$\mathcal{N}$ argument.
The fourth order tunneling terms are of the form $\text{tr}(\hat{G}_{0}\hat{T}\hat{G}_{0}\hat{T}\hat{G}_{0}\hat{T}\hat{G}_{0}\hat{T}).$
These processes involves two or three dots. Consider for example the
three dot term (with consecutive dots labeled $i,j,k$),
\begin{align*}
\text{tr}(\hat{G}_{0}\hat{T}\hat{G}_{0}\hat{T}\hat{G}_{0}\hat{T}\hat{G}_{0}\hat{T}) & =\sum_{\stackrel{ijk}{\alpha_{1,\ldots,}\alpha_{4}}}(\hat{G}_{0})_{i,\alpha_{1}}\hat{T}_{\alpha_{1}\alpha_{2}}^{ij}(\hat{G}_{0})_{j,\alpha_{2}}\hat{T}_{\alpha_{2}\alpha_{3}}^{jk}(\hat{G}_{0})_{k,\alpha_{3}}\hat{T}_{\alpha_{3}\alpha_{4}}^{kj}(\hat{G}_{0})_{j,\alpha_{4}}\hat{T}_{\alpha_{4}\alpha_{1}}^{ji}.
\end{align*}
 Now the tunneling amplitudes $\tilde{t}$ are of the form $\tilde{t}_{\alpha\beta}^{ij}=|\tilde{t}|e^{i\chi_{\alpha\beta}^{ij}},$
where $\chi_{\alpha\beta}^{ij}$ is a phase associated with the link
$ij$ and energy levels $\alpha,\beta.$ The key point is that for
irregular dots, the phases $\chi_{\alpha\beta}^{ij}$ are random.
For the case of a large number of levels, the random phases cause
the vanishing of all terms except for the case $\alpha_{4}=\alpha_{2}$
where the random phases cancel exactly. Thus there are only three
independent energy indices to be summed over resulting in a factor
of $\mathcal{N}^{3}.$ However since the $\tilde{t}$ scale as $1/\mathcal{N},$
it is evident that the overall scaling of this term is $1/\mathcal{N}.$
In general, the number of independent energy indices in the perturbative
expansion of the tunneling action equals the number of dots involved
in that term. 

We now discuss the case of ballistic point contacts. The fourth order
three-dot term can be written as 
\begin{align*}
\text{tr}(\hat{G}_{0}\hat{T}\hat{G}_{0}\hat{T}\hat{G}_{0}\hat{T}\hat{G}_{0}\hat{T}) & =\sum_{\stackrel{ijl,\mathbf{k}_{\perp}}{k_{1,\ldots,}k_{4}}}(\hat{G}_{0})_{i,k_{1}}\hat{T}_{k_{1}k_{2}}^{ij}(\hat{G}_{0})_{j,k_{2}}\hat{T}_{k_{2}k_{3}}^{jl}(\hat{G}_{0})_{l,k_{3}}\hat{T}_{k_{3}k_{4}}^{lj}(\hat{G}_{0})_{j,k_{4}}\hat{T}_{k_{4}k_{1}}^{ji},
\end{align*}
where $k_{1},\ldots,k_{4}$ are longitudinal momenta and we have suppressed
the transverse momentum label $\mathbf{k}_{\perp}$ for brevity. Since
the tunneling elements scale as $|\tilde{t}|\sim1/(\sqrt{N_{\text{ch}}}\mathcal{N}_{1D}),$
each term in the above sum scales as $1/(N_{\text{ch}}^{2}\mathcal{N}_{1D}^{4}).$
Now the sum over the four longitudinal momenta brings a factor of
$\mathcal{N}_{1D}^{4},$ and the sum over the transverse momentum
gives a factor $N_{\text{ch}}.$ Thus we find that the above fourth
order contribution scales as $1/N_{\text{ch}}.$ In order to be able
to neglect this fourth order term, we require $N_{\text{ch}}\gg1,$i.e.,
the width of the point contact should be much larger than the Fermi
wavelength.

There is a second very important consequence of large-$\mathcal{N}$
that provides a crucial simplification in nonequilibrium situations
and which has not been appreciated in the literature. This relates
to the temporal variation of the the $F_{k}$ under general nonequilibrium
conditions. It is convenient to work with the Wigner representation,
$F_{k}(t,t')\equiv\int(d\epsilon)F_{k}(\epsilon,\tau)e^{-i\epsilon(t-t')},$
where $\tau=(t+t')/2,$ and the relation with the time-dependent distribution
function is $F_{k}(\epsilon,t)=1-2f_{k}(\epsilon,t).$ The total number
of electrons in the $k^{\text{th}}$ dot is $N_{0}+n_{k}^{c}(t)=\int d\epsilon\,\nu(\epsilon)f(\epsilon,t),$
where $n_{k}^{c}(t)$ is the classical component of the number field
conjugate to the quantum component of the phase, $\phi_{k}^{q}.$
In the rest of the paper, we will be specifically interested in the
case of constant $N_{0}.$ More general, time-dependent $N_{0}$ can
if a time-dependent gate voltage is applied to the quantum dots. Thus
in our case we have 
\begin{align}
\frac{dn_{k}^{c}}{dt} & =V\int d\epsilon\,\nu(\epsilon)\frac{df_{k}(\epsilon,t)}{dt}.\label{eq:ncl-change}
\end{align}
The RHS of Eq. (\ref{eq:ncl-change}) is, by using the continuity
equation, simply the net current into the dot, and is given by the
functional derivative $\langle\delta S/\delta\phi_{k}^{q}(t)\rangle_{\phi},$
which has the form $g\int d\epsilon\,h(\epsilon,t)\equiv j_{k-1,k}(t)-j_{k,k+1}(t).$
Consequently, the continuity equation leads us to a kinetic equation
for the distribution $f_{k}(\epsilon,t)$ of the form $V\nu(\mu)df_{k}/dt+gh(\epsilon,t)=0.$
The quantity $h$ is a functional of the distributions $\{f_{k}\}$
and also depends on the tunneling conductance and electric field.
Recognizing $V\nu(\mu)=1/\delta,$ we find that the distribution function
evolves with a large characteristic time scale that is proportional
to $1/g\delta$ and increases linearly with the total number of electrons
in the grain ($\delta\sim1/\mathcal{N}$). We now assume that the
grains are coupled to an external thermal bath, whose effect we model
by an additional relaxation term in the kinetic equation, i.e.,
\begin{align}
\frac{df_{k}}{dt} & =-g\delta h[f]+\frac{f_{k}-f_{k}^{\text{eq}}}{\tau_{eb}},\label{eq:kinetic}
\end{align}
where $f_{k}^{\text{eq}}$ is the equilibrium Fermi-Dirac distribution
function and $\tau_{eb}$ is the electron-bath relaxation time. If
$1/\tau_{eb}\ll g\delta,$ then the distribution functions $f_{k}$
may be approximated by their equilibrium values. We will now proceed
with this, and hence $F_{k}(\epsilon)=\tanh(\epsilon/2T).$ In contrast,
in the usual Hubbard models, the electron distribution function at
every site is a time dependent quantity under general nonequilibrium
conditions since in that case there is no large-$\mathcal{N}$ mitigating
factor.

\subsection{Keldysh-AES action}

We resume our derivation of the effective Keldysh AES action. Henceforth
we will describe tunneling in both the granular metal as well as the
point contact cases by the action in Eq. (\ref{eq:Stun-2}) and note
that $g$ can have different forms for the two cases. Now let us manipulate
$S_{\mbox{tun}}$ to a more dealable form. We introduce new fields
$C$ and $S$ defined as 
\begin{eqnarray}
C=\exp(\iota\phi^{c})\cos\left(\frac{\phi_{q}}{2}\right) & \mbox{ , } & S=\exp(\iota\phi^{c})\sin\left(\frac{\phi_{q}}{2}\right).
\end{eqnarray}
These are related to the $\hat{\phi}$ fields in Eq. (\ref{eq:hat-fields})
through 
\begin{eqnarray}
\exp(\iota\hat{\phi})=C+\iota S\sigma_{1} & \mbox{ , } & \exp(-\iota\hat{\phi})=\bar{C}-\iota\bar{S}\sigma_{1}.
\end{eqnarray}
The tunneling action under equilibrium conditions then takes the form
\begin{equation}
S_{\mbox{tun}}=4g\sum_{k}\int_{t,t^{'}}\begin{bmatrix}\bar{C}_{k,1} & -\iota\bar{S}_{k,1}\end{bmatrix}_{t}\begin{bmatrix}0 & \Sigma_{k,1}^{A}\\
\Sigma_{k,1}^{R} & \Sigma_{k,1}^{K}
\end{bmatrix}_{t-t^{'}}\begin{bmatrix}C_{k,1}\\
\iota S_{k,1}
\end{bmatrix}_{t^{'}},
\end{equation}
where 
\begin{eqnarray}
\Sigma_{k,1}^{R(A)}(t) & = & \iota\left(G^{R(A)}(t)G_{k}^{K}(-t)+G_{k+1}^{K}(t)G^{A(R)}(-t)\right),\label{eq:sigma-causal}\\
\Sigma_{k,1}^{K}(t) & = & \iota\left(G_{k}^{K}(-t)G_{k+1}^{K}(t)-(G^{R}-G^{A})_{t}(G^{R}-G^{A})_{-t}\right),
\end{eqnarray}
with $G_{k}^{K}=F_{k}(G^{R}-G^{A})$. It is evident from Eq. (\ref{eq:sigma-causal})
that $\Sigma^{R(A)}$ also have a causal structure, i.e., $\Sigma^{R}(t)\propto\Theta(t)$
etc. Under general nonequilibrium conditions, the quantities $\Sigma^{R,A,K}(t,t')$
describing particle-hole excitations in the dots depend on both the
time arguments, and not just their difference. 

Let $F_{b}(\epsilon)=\coth(\epsilon/2T)=1+2f_{b},$ where $f_{b}$
is the equilibrium Bose distribution function. We make use of the
following identities, 
\begin{eqnarray}
\left(G^{R}-G^{A}\right)_{\epsilon} & = & -\iota,\\
\int_{\epsilon}\frac{1}{2\pi}\left(F(\epsilon+\omega)-F(\epsilon)\right) & = & \frac{\omega}{\pi},\\
\int_{\epsilon}\frac{1}{2\pi}\left(1-F(\epsilon-\omega)F(\epsilon)\right) & = & \frac{\omega}{\pi}F_{b}(\omega).
\end{eqnarray}
to obtain, 
\begin{eqnarray}
\left(\Sigma_{k,1}^{R}-\Sigma_{k,1}^{A}\right)_{\omega}=\iota\int_{\epsilon}\frac{1}{2\pi}\left(F_{k+1}(\epsilon)-F_{k}(\epsilon-\omega)\right)=\frac{\iota}{\pi}\omega,\label{id1}\\
(\Sigma_{k,1}^{K})_{\omega}=\iota\int_{\epsilon}\frac{1}{2\pi}\left(1-F_{k+1}(\epsilon)F_{k}(\epsilon-\omega)\right)=\frac{\iota}{\pi}\omega F_{b}(\omega).\label{id2}
\end{eqnarray}
We will later find it convenient to work in the $\pm$ Keldysh contour.
Hence we re-express our phase action in this contour. We ignore $N_{0}$
by assuming that it can be set to zero by some gate voltage. We have,
\begin{eqnarray}
S_{C}[n,\phi] & = & \frac{1}{E_{C}}\sum_{k}\int_{t}[(\partial_{t}\phi_{k}^{+})^{2}-(\partial_{t}\phi_{k}^{-})^{2}],\\
S_{\mbox{tun}}[\phi] & = & g\sum_{k}\int_{t.t^{'}}\begin{pmatrix}\exp(-\iota\phi_{k,1}^{+}) & \exp(-\iota\phi_{k,1}^{-})\end{pmatrix}_{t}L_{k,1}(t-t^{'})\begin{pmatrix}\exp(\iota\phi_{k,1}^{+})\\
\exp(\iota\phi_{k,1}^{-})
\end{pmatrix}_{t^{'}},\label{eq:stun-pm}\\
L & = & \frac{1}{4}\begin{pmatrix}\Sigma^{R}+\Sigma^{A}+\Sigma^{K} & \Sigma^{R}-\Sigma^{A}-\Sigma^{K}\\
-\Sigma^{R}+\Sigma^{A}-\Sigma^{K} & -\Sigma^{R}-\Sigma^{A}+\Sigma^{K}
\end{pmatrix}.\label{Lexp}
\end{eqnarray}
Note that the diagonal elements of the matrix $L$ written in the
$\pm$ basis contain the combination $\Sigma^{R}+\Sigma^{A}$ and
the off-diagonal elements contain $\Sigma^{R}-\Sigma^{A}.$ In the
(equilibrium) Matsubara formalism, finite winding numbers of the phase
fields must be considered to bring out the charge quantization effects.
In our continuous time formalism, the charge quantization effects
are brought out by a procedure discussed, for example, in Ref. \cite{altland2010condensed}
that we briefly describe below.

\subsection{Phase windings and charge quantization}

We are interested in the small tunneling regime, $g\lesssim1.$ In
this regime, the phases in each dot fluctuate strongly and hence we
represent the action in terms of the conjugate variables, i.e., the
number fields. For this, we first perform a Hubbard-Stratanovich decoupling
of the charging term, which leads to the following action in the phase-charge
representation: 
\begin{equation}
S[n,\phi]=\sum_{k}\int_{t}\left([n_{k}^{c}+N_{0}]\partial_{t}\phi_{k}^{q}+n_{k}^{q}\partial_{t}\phi_{k}^{c}-2E_{C}n_{k}^{c}n_{k}^{q}\right)+S_{\mbox{tun}}[\phi].\label{eq:cl-q-action}
\end{equation}
To properly understand the quantization of the charge degrees of freedom,
we first work in a contour, $t\in[0,P]$. The requirement that $\phi^{-}(0)=\phi^{+}(0)+2\pi W$
($W$ is an integer) leads us to an unconstrained field, $\phi^{c},$
and, 
\begin{equation}
\phi_{q}(t)=\tilde{\phi}_{q}(t)+\frac{2\pi W}{P}(t-P),\label{eq:phi-q}
\end{equation}
with Dirichlet conditions, $\tilde{\phi}^{q}(0)=\tilde{\phi}^{q}(P)=0$.
Consider first the situation where tunneling is absent. Using Eq.
(\ref{eq:phi-q}) in the first term of Eq. (\ref{eq:cl-q-action}),
we see that the partition function has contributions of the form $\sum_{W}e^{\iota2\pi(n^{c}+N_{0})W},$which
vanishes unless $n^{c}+N_{0}$ is an integer. Writing $N_{0}=[N_{0}]+n_{g},$
where $[N_{0}]$ is the integer part of $N_{0}$ and $n_{g}\in[0,1)$
is the residual ``gate charge'' on a dot, the integration over the
Hubbard-Stratonovich field $n^{c}$ is equivalent to a sum over integers,
$\sum_{[n^{c}]-n_{g}},$ where $[n^{c}]$ is the integer part of $n^{c}.$
Making a change of variables, $n^{c}\rightarrow n^{c}-n_{g},$ the
sum becomes one over integer values of $n^{c}.$ Now the part of the
action containing the time derivative of the classical phase field
is a function only of the boundary values of the field. Performing
the path integral over the boundary fields gives us the constraint
that $n^{q}=0$ at the boundaries. Let's now imagine turning on the
tunneling at some time. From the structure of the tunneling action,
Eq. (\ref{eq:stun-pm}), it is clear that $n^{+}$ and $n^{-}$ can
change only in integer steps. This quantization condition is independent
of the time boundary or the length of the time interval. Translated
back in the language of the Keldysh closed-time contour, the condition
that the \emph{initial }values of $n^{c}$ can only take integer values
together the fact that boundary values of $n^{q}$ are zero, one concludes
that $n^{+}(-\infty)=n^{-}(-\infty)\in Z,$ and both change in only
in integer steps during tunneling events. In this paper, we are interested
in the Mott insulator regime with zero gate charge, i.e., $n_{g}=0$
(or integer $N_{0}$) and therefore we drop the $N_{0}\partial_{t}\phi^{q}$
term in the action. The point $n_{g}=1/2$ is special due to degeneracy
between $n^{c}=0,1.$ The gate charge, $n_{g},$ can also be made
to fluctuate in time by using a time-dependent gate voltage. These
different scenarios can also be studied using our formalism and will
be taken up elsewhere.

\subsection{Functional representation of charge current}

Here we obtain the functional representation for the charge current
in the presence of a constant electric field. The electric field is
introduced in the form of a time-dependent vector potential that is
turned on at some instant of time, say $t=0.$ In every link, the
classical component of the vector potential has the form 
\begin{align}
A_{k,1}^{c}(t) & =\Theta(t)Dt,\label{eq:Acl}
\end{align}
where $D$ is the potential energy change across a link as already
mentioned in Sec. \ref{sec:Introduction}. This changes the tunneling
part of the action by incorporating the Peierls shifts in the phase
differences, $\phi_{k,1}^{c,q}(t)\rightarrow\phi_{k,1}^{c,q}(t)+A_{k,1}^{c,q}(t).$
The tunneling part of the action now has the form

\begin{equation}
S_{\mbox{tun}}[\phi,A^{c},A^{q}]=g\sum_{k}\int_{t,t'}\begin{bmatrix}(e_{k,1}^{+}(t))^{*} & (e_{k,1}^{-}(t))^{*}\end{bmatrix}L(t-t')\begin{bmatrix}e_{k,1}^{+}(t')\\
e_{k,1}^{-}(t')
\end{bmatrix},
\end{equation}
where, $e_{k,1}^{\pm}(t)=\exp(\iota\phi_{k,1}^{\pm}(t)-\iota A_{k,1}^{\pm}(t))$.
The functional representation of the classical component of the charge
current in a link, $\hat{J}_{k,1}[A^{c}(t)],$ is obtained by taking
the functional derivative with respect to $A_{k,1}^{q}(t),$ and setting
this quantum source term to zero:
\begin{align}
\hat{J}_{k,1}(\tau) & =-\iota g\int_{t}\left[(e_{\tau}^{+})^{*}L_{\tau t}^{++}e_{t}^{+}-(e_{t}^{+})^{*}L_{t\tau}^{++}e_{\tau}^{+}+(e_{\tau}^{+})^{*}L_{\tau t}^{+-}e_{t}^{-}+(e_{t}^{+})^{*}L_{t\tau}^{+-}e_{\tau}^{-}\right.\nonumber \\
 & \qquad\left.-(e_{\tau}^{-})^{*}L_{\tau t}^{-+}e_{t}^{+}-(e_{t}^{-})^{*}L_{t\tau}^{-+}e_{\tau}^{+}-(e_{\tau}^{-})^{*}L_{\tau t}^{--}e_{t}^{-}+(e_{t}^{-})^{*}L_{t\tau}^{--}e_{\tau}^{-}\right].\label{eq:J-func}
\end{align}
Here we have suppressed the site indices and written the time arguments
as subscripts for brevity.

\section{Transient current response \label{sec:Transient-current-response}}

In this Section, we obtain the current response to leading order (in
$g$) upon turning on the uniform electric field by performing the
average of the current functional in Eq. (\ref{eq:J-func}) over the
phase fields. This primarily involves a calculation of the bond correlators
defined as 
\begin{equation}
\Pi_{\sigma\sigma'}(\tau,\tau')=\left<\exp\left[-\iota\phi_{j,1}^{\sigma}(\tau)+\iota\phi_{j,1}^{\sigma'}(\tau')\right]\right>.\label{bondcor}
\end{equation}
Here $\left<...\right>$ denotes averaging with the full action, $S[n,\phi]$. 

We calculate the bond correlators as a perturbation series in the
tunneling conductance $g$, by treating the charging action as the
bare action and expanding the tunneling part in the exponential to
various orders in $g$. We denote $\left<...\right>_{0}$ to represent
averaging with the bare action. The bare bond correlator, $\Pi_{\sigma\sigma'}^{(0)}$
factorizes into a product of two single site correlators, 
\begin{align}
\Pi_{\sigma\sigma'}^{(0)}(\tau,\tau') & =C_{\sigma\sigma'}(\tau,\tau')C_{\sigma'\sigma}(\tau',\tau),\label{eq:pi0}
\end{align}
where 
\begin{align}
C_{\sigma\sigma'}(\tau,\tau') & =\left\langle e^{-\iota(\phi^{\sigma}(\tau)-\phi^{\sigma'}(\tau'))}\right\rangle _{0}.\label{eq:C-def}
\end{align}
Let us first consider $C_{+-}(\tau-\tau')$. Performing the functional
integral over the phase fields $\phi^{\pm}$ we get the equations,
\begin{equation}
\partial_{t}n^{+}=-\delta(t-\tau)\mbox{ , }\partial_{t}n^{-}=-\delta(t-\tau^{'}).\label{nchange}
\end{equation}
The solution depends on the boundary conditions at $t=-\infty.$ We
assume that in the remote past, the system is in thermal equilibrium,
and hence the probability distribution for $n^{c}$ is $P(n^{c})=\exp(-\beta(n^{c})^{2}E_{C})/\sum_{n=-\infty}^{\infty}\exp(-\beta E_{C}n^{2}).$
In the zero temperature limit, $P(n^{c})=\delta_{n^{c},0}$. Furthermore
since $n^{q}(-\infty)=0,$ we have $n^{+}(-\infty)=n^{-}(-\infty)=0.$
Thus the solution to Eq. (\ref{nchange}) is 
\begin{equation}
n^{+}(t)=-\Theta(t-\tau)\mbox{ , }n^{-}(t)=-\Theta(t-\tau^{'}).
\end{equation}
Plugging this back, we get, 
\begin{equation}
C_{+-}(\tau,\tau^{'})=\exp(\iota E_{C}(\tau-\tau^{'})).
\end{equation}
Similarly, 
\begin{eqnarray}
C_{-+}(\tau,\tau^{'}) & = & \exp(-\iota E_{C}(\tau-\tau^{'})),\\
C_{\pm\pm}(\tau,\tau^{'}) & = & \exp(\mp\iota E_{C}|\tau-\tau^{'}|).
\end{eqnarray}
Using these site correlators in Eq. (\ref{bondcor}) for the bond
correlators in Eq. (\ref{eq:J-func}), and using the causal structure
of $\Sigma^{R(A)},$ we obtain the following expression for the leading
order nonequilibrium current
\begin{align}
J(\tau) & =\frac{g}{2\pi}\int_{-\infty}^{\tau}\mbox{d}t\left[e^{\iota D(\tau\Theta(\tau)-t\Theta(t))}\left\{ 2\Sigma^{R}(\tau-t)\cos(2E_{C}(\tau-t))\right.\right.\nonumber \\
 & \qquad\left.\left.-2\iota\Sigma^{R}(\tau-t)\cos(2E_{C}(\tau-t))\right\} +\text{c.c.}\right].\label{eq:J-transient1}
\end{align}
Since the upper limit of the integral is $t=\tau$ and $\Sigma^{A}(t)$
has a $\Theta(-t)$ structure, we can replace 
\begin{equation}
\Sigma^{R}(\tau-t)\rightarrow\Sigma^{R}(\tau-t)-\Sigma^{A}(\tau-t),
\end{equation}
and use the relation for the Fourier transform, Eq. (\ref{id1}).
For $\tau<0$, the average current clearly vanishes. Let us split
the integral in Eq. (\ref{eq:J-transient1}) into two parts, $J=J_{<}+J_{>}$,
where $J_{<}$ involves integration from $t=-\infty$ to 0 and in
$J_{>},$ $t=0$ to $\tau:$
\begin{align}
J_{<}(\tau) & =\frac{ge^{\iota D\tau}\Theta(\tau)}{(2\pi)^{2}}\int_{-\infty}^{\infty}\mbox{d}\omega\int_{-\infty}^{0}\mbox{d}t\,\left[e^{\iota(2E_{C}-\omega)(\tau-t)}(\omega-|\omega|)\right.\nonumber \\
 & \qquad\qquad\qquad\qquad\qquad\qquad\qquad\qquad\left.+e^{-\iota(2E_{C}+\omega)(\tau-t)}(\omega+|\omega|)+\text{c.c.}\right],\nonumber \\
J_{>}(\tau) & =\frac{g\Theta(\tau)}{(2\pi)^{2}}\int_{-\infty}^{\infty}\mbox{d}\omega\int_{0}^{\tau}\mbox{d}t\,\left[e^{\iota(2E_{C}-\omega+D)(\tau-t)}(\omega-|\omega|)\right.\nonumber \\
 & \qquad\qquad\qquad\qquad\qquad\qquad\qquad\left.+e^{-\iota(2E_{C}+\omega-D)(\tau-t)}(\omega+|\omega|)+\text{c.c.}\right].\label{eq:J<J>}
\end{align}
 After performing the time integration and some simple manipulations,
we get
\begin{align}
J_{<}(\tau) & =-\frac{4\iota ge^{\iota D\tau}\Theta(\tau)}{(2\pi)^{2}}\int_{0}^{\infty}d\omega\frac{\omega\cos((\omega+2E_{C})\tau)}{\omega+2E_{C}}+\text{c.c.},\nonumber \\
J_{>}(\tau) & =\frac{2\iota g\Theta(\tau)}{(2\pi)^{2}}\left[\int_{0}^{\infty}d\omega\frac{\omega e^{\iota(\omega+2E_{C}+D)\tau}}{\omega+2E_{C}+D}+\int_{0}^{\infty}d\omega\frac{\omega e^{-\iota(\omega+2E_{C}-D)\tau}}{\omega+2E_{C}-D}\right.\nonumber \\
 & \qquad\qquad\qquad\qquad\qquad\left.-\int_{0}^{\infty}d\omega\frac{\omega(4E_{C}+2\omega)}{(2E_{C}+\omega)^{2}-D^{2}}\right]+\text{c.c.}.\label{eq:J<J>2}
\end{align}
Now, using 
\begin{eqnarray}
\int_{0}^{\infty}d\omega\frac{\omega e^{\iota\omega\tau}}{\omega+x} & = & \frac{\iota}{\tau}-x\int_{x}^{\infty}du\frac{e^{\iota(u-x)\tau}}{u},\nonumber \\
 & = & \frac{\iota}{\tau}-xe^{-\iota x\tau}(\iota\pi\Theta(x)-\ei(\iota x\tau)),
\end{eqnarray}
the expression for the current simplifies to 
\begin{align}
J(\tau) & =\frac{2\iota g\Theta(\tau)}{(2\pi)^{2}}\left[(2E_{C}+D)(\ei(\iota(2E_{C}+D)\tau)-\ei(-\iota(2E_{C}+D)\tau))\right.\nonumber \\
 & \qquad-(2E_{C}-D)(\ei(\iota(2E_{C}-D)\tau)-\ei(-\iota(2E_{C}-D)\tau))\nonumber \\
 & \qquad-2\iota E_{C}\sin(D\tau)(\ei(\iota2E_{C}\tau)+\ei(-\iota2E_{C}\tau))\nonumber \\
 & \qquad\left.-2\pi\iota(2E_{C}+D)\Theta(2E_{C}+D)+2\pi\iota(2E_{C}-D)\Theta(2E_{C}-D)\right].\label{eq:J-trans2}
\end{align}
The current response at long times $\tau\gg\tau_{+}=\text{max}[|D+2E_{C}|^{-1},|D-2E_{C}|^{-1}]$
has two components ($J(\tau\gg\tau_{0})=J_{\text{dc}}+J_{\text{tr}}$):
a dc part, 
\begin{align}
J_{\text{dc}} & =\frac{g\Theta(\tau)}{\pi}\left[(D-2E_{C})\Theta(D-2E_{C})+(D+2E_{C})\Theta(-2E_{C}-D)\right]\label{eq:dc-response}
\end{align}
and a transient part, 
\begin{align}
J_{\text{tr}} & \approx-\frac{4g\Theta(\tau)}{(2\pi)^{2}E_{C}}\frac{1}{\tau^{2}}\left[\left(\frac{D-2E_{C}}{D+2E_{C}}\right)\sin((D+2E_{C})\tau)+\left(\frac{D+2E_{C}}{D-2E_{C}}\right)\sin((D-2E_{C})\tau)\right],\label{eq:Jtrans-final}
\end{align}
that oscillates with the two beat frequencies $\omega_{\pm}=|D\pm2E_{C}|,$
and slowly decays in accordance with an inverse square law in time.
Such oscillations are absent in classical $RC$ networks subjected
to a constant electric field, where only exponential relaxation may
occur. 
The amplitudes of the two oscillation frequencies are inversely
related. Close to a resonance, $D=\pm2E_{C},$ the amplitude of the
faster mode tends to vanish and the slower mode dominates. At high
fields, $|D|\gg2E_{C},$ the beat frequencies are approximately $\omega_{\pm}\approx|D|=\omega_{B},$
where $\omega_{B}$ is the Bloch oscillation frequency for noninteracting
electrons. It is instructive to compare with the fermionic Hubbard
chain at half-filling - a quantum model that is the dissipation-free
counterpart of ours. At strong electric fields, the Bloch oscillations
in this model also occur \cite{eckstein2010dielectric} at $\omega_{B},$
and which has a simple physical explanation. Consider a noninteracting
model of fermions hopping on a one-dimensional lattice: 

\begin{align}
H_{\text{el}}^{(0)} & =-t\sum_{\langle ij\rangle\sigma}[c_{i\sigma}^{\dagger}c_{j\sigma}+\text{h.c.}]+\sum_{i\sigma}\epsilon_{i}n_{i\sigma},\label{eq:H-nonint}
\end{align}
where $\epsilon_{j}=Dj$ is the linearly varying potential energy
in the presence of a constant electric field. As is well-known {[}see
eg. \cite{emin1987phonon}{]}, the above Hamiltonian is easily diagonalized
by the transformation 
\begin{align}
f_{n} & =\sum_{i}J_{i-n}(2t/D)c_{i},\label{eq:ws-states}
\end{align}
which gives us a discrete spectrum, the Wannier-Stark ladder, with
energies $E_{n}=nD,$ with $n$ an integer. The wavefunction corresponding
to $E_{n}$ is localized, centered around the site $n,$ and with
a spatial extent of the order of $L=2t/D.$ Since there is no matrix
element connecting different Wannier-Stark levels, no net current
flows in the system. If the gain in potential energy across a link,
$D,$ greatly exceeds the tight binding hopping energy, then the Wannier-Stark
states are highly localized. Introducing now a small local Hubbard
repulsion term of strength $E_{C}$ in Eq. (\ref{eq:H-nonint}), we
find that the interaction remains approximately local even in the
Wannier-Stark basis. For $D\gg E_{C},$ the energy levels are approximately
$nD,$ which leads to Bloch oscillations at frequency $\omega_{B}.$

    \begin{figure}
    \begin{center}$
    \begin{array}{ccc}
    \includegraphics[width=.44\linewidth]{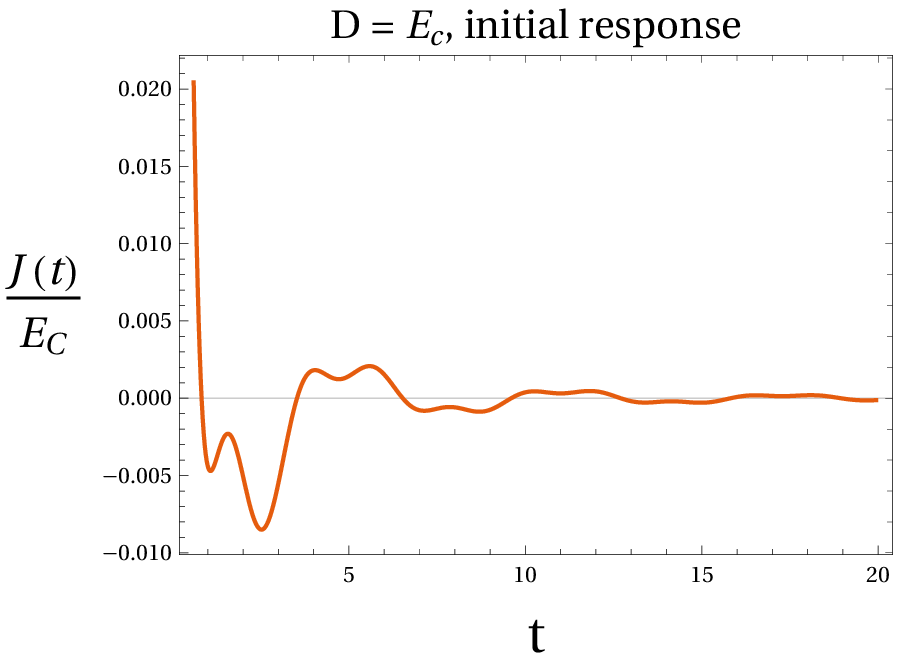}&
    \hspace*{.5cm}&
    \includegraphics[width=.45\linewidth]{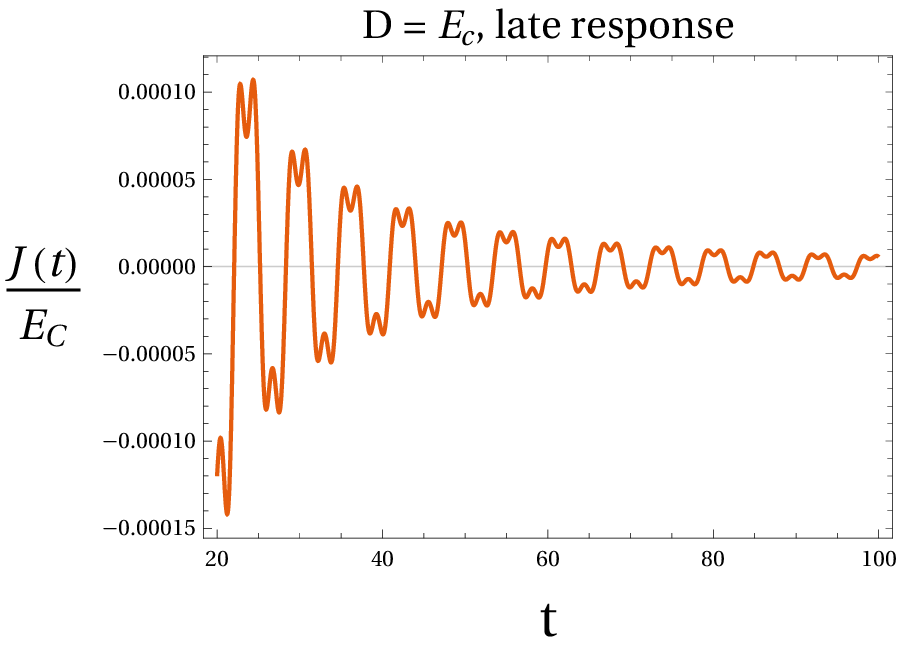}
    \end{array}$
    
    $
    \begin{array}{ccc}
    \includegraphics[width=.44\linewidth]{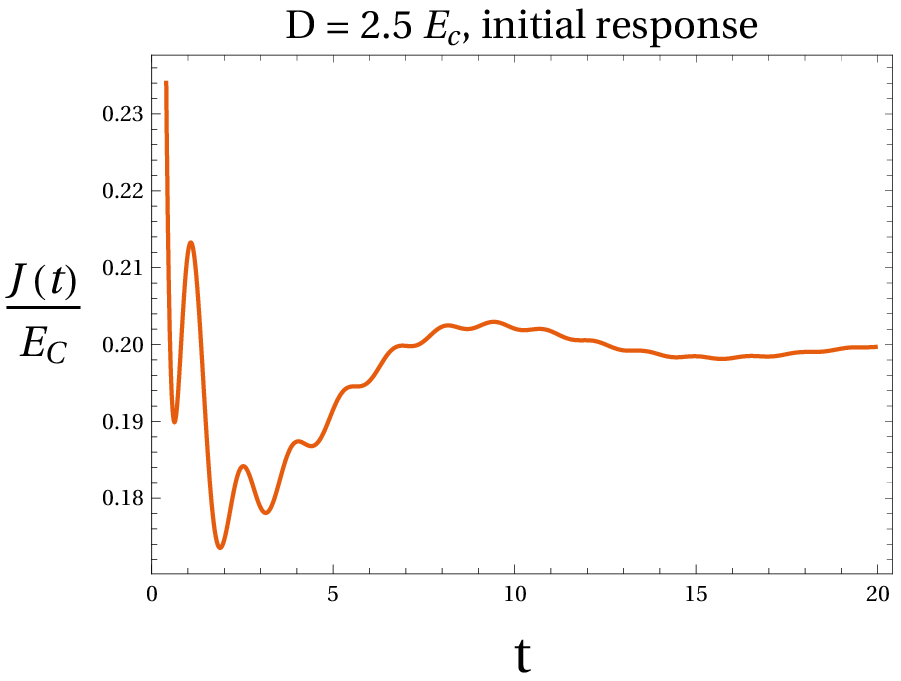}&
    \hspace*{.5cm}&
    \includegraphics[width=.45\linewidth]{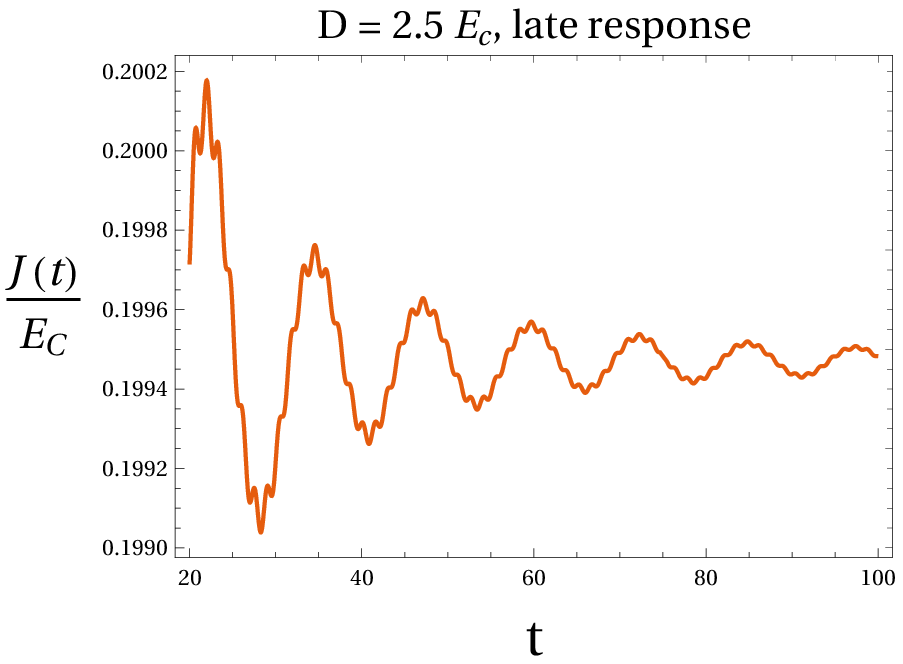}
    \end{array}$
    
     $
    \begin{array}{ccc}
    \includegraphics[width=.44\linewidth]{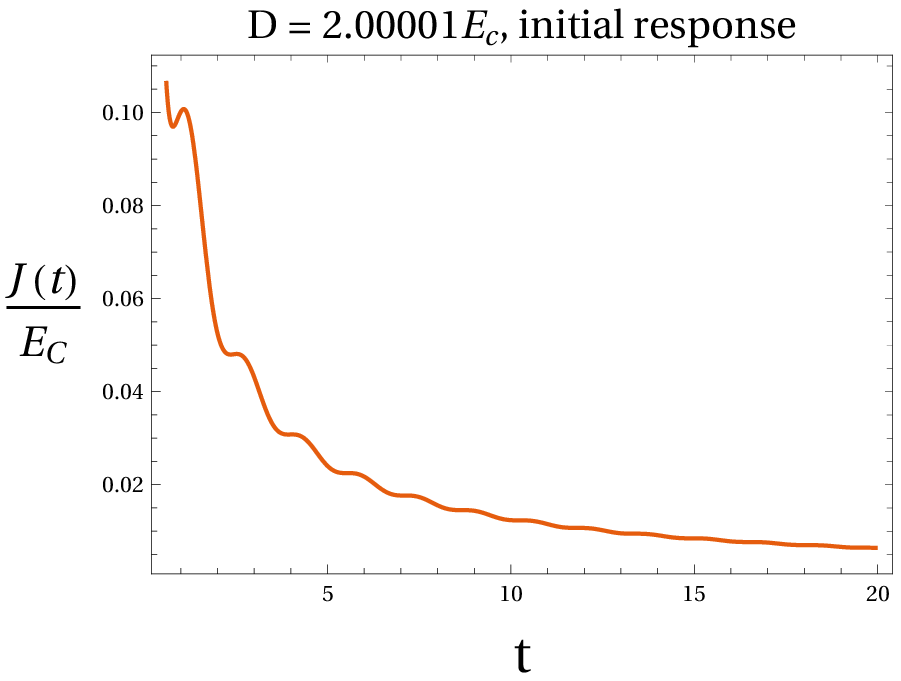}&
    \hspace*{.5cm}&
    \includegraphics[width=.45\linewidth]{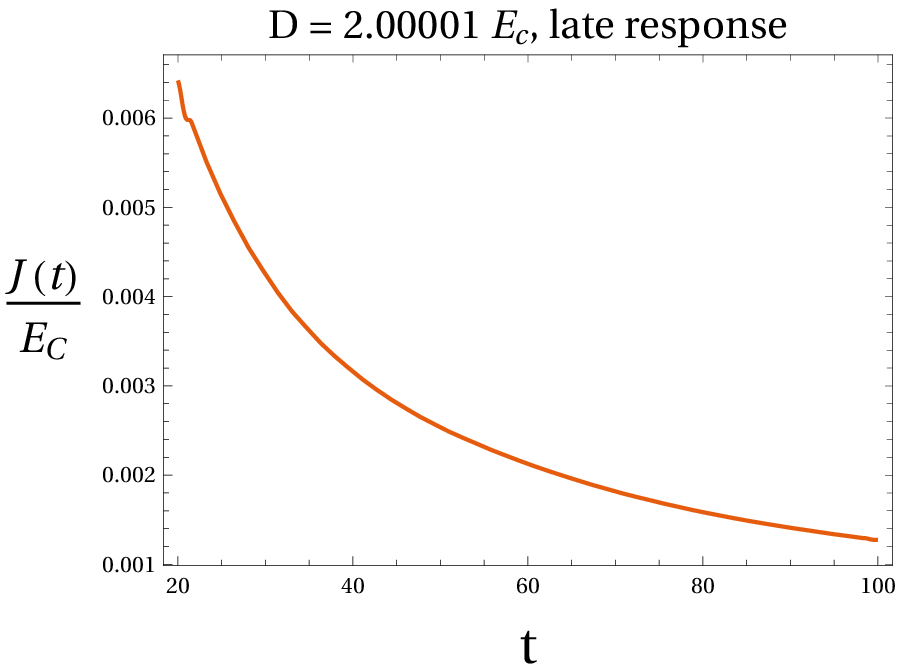}
    \end{array}$

    \end{center}
    \caption{\label{fig:transient} The current response to leading order in $g$ after an electric field is turned on as described by Eq. \ref{eq:J-trans2}. The plots to the left show 
the initial time reponse and those to the right show the late time response where the power law decay of the oscillatory behavior is seen.
 The effect of correlations in the late time response is seen in the form of beating frequencies. A finite steady state DC response exists only for $D>2E_{C}$}
    
    \end{figure}

Consider now the short-time current response. Above the threshold
field, $D_{T}>2E_{C},$ a finite dc response exists unlike the dissipationless
Hubbard chain at half filling. However, the Bloch-like oscillations
are present both above and below the threshold field. At short times
$\tau\ll\tau_{-}=\text{min}[|D+2E_{C}|^{-1},|D-2E_{C}|^{-1}],$ the
current response is 
\begin{align}
J(\tau) & \approx\frac{g\Theta(\tau)}{\pi}D-\frac{8g\Theta(\tau)}{(2\pi)^{2}}D(2E_{C}\tau)\left[\ln(1/2E_{C}\tau)+2-\gamma\right],\label{eq:j-small_time}
\end{align}
where $\gamma\approx0.577$ is the Euler-Mascheroni constant. Remarkably,
the initial current response, $J=gD/\pi$ is independent of the charging
energy, $E_{C},$ and appears to be physically related to the fact
that sudden changes in the potential effectively short-circuit a capacitor.
Plots of the current response for different applied electric field strengths are given in Fig. \ref{fig:transient}
The transient current response is a central result of this paper.
 
\subsection{Long time response: the effect of dissipation} 
The $1/\tau^{2}$ decay of the amplitude of current oscillations arises from the ohmic dissipation kernel ($\Sigma^{K}(\omega)\sim |\omega| $ in the zero temperature limit which implies $\Sigma^{K}(t)\sim 1/t^{2}$). Higher order corrections (in $g$) will similarly decay as $1/\tau^{2n}$, with $n>1.$ Thus for small $g$, one clearly expects that at long times, the transient part of the current response will be dominated by the leading order term and hence a $1/\tau^{2}$ decay of the oscillatory response. The dc part of the current response, on the other hand, is not neccessarily dominated by the leading order in $g$ term. For instance in our case, the leading order dc response vanishes for $D<2E_{c}$. However we will show in the following section that a finite dc current exists even for small values of $D,$ and is dominated by higer order in $g$ terms.

We now argue that the Bloch-like oscillations and power-law decay of the transient response, a manifestation of charge quantization, are not expected to hold for arbitrary long times. Physically, the finite dc response is a consequence of dissipation, which in turn, should ultimately introduce a time scale beyond which an exponential decay rather than a power-law decay should occur. To resolve this, we examine the validity of approximating the ohmic kernel by its zero temperature limit.

The existence of a finite dc current component in response to a dc driving field (see Sec. \ref{sec:Steady-state} below) implies a finite power dissipation, $W \sim J_{\mbox{dc}}D,$ where $J_{\mbox{dc}}$ is the dc current. The coupling to an external heat bath is necessary for a steady dc response, for it ensures that the electron distribution in a dot does not run off to infinite temperature as a result of this Joule heating. We assumed earlier that the coupling with the external bath is weak in the sense that the energy relaxation time with the bath, $\tau _{eb}$ is much greater than the typical electron energy relaxation time within a grain, $\tau_{R} \sim \delta/U^2 \sim  O(1/\mathcal{N}),$ where $U$ is the matrix element for electron-electron coupling in the grain. The separation of these time scales makes the electron distribution thermal even when $W$ is finite. The excess thermal energy in a grain is $W \tau_{eb},$ and this is shared by the $\mathcal{N}$ electrons in the grain, implying a finite temperature $T_{D} \sim J_{\mbox{dc}}D\tau_{eb}/\mathcal{N}.$ The electron-bath relazation time $\tau_{eb}$ will generally decrease with increasing $\mathcal{N},$ and we expect $\tau_{eb}\sim \mathcal{N}^{-2/3}$ if we assume the bath degrees of freedom essentially interact with the surface of the grain. Further, if $\tau_{eb}$ is due to electron-phonon coupling, it may also have a temperature dependence; i.e., 
\begin{equation}
\tau_{eb} \sim (1/\mathcal{N})^{2/3} (1/T_D)^{n},
\end{equation}
where $n \geq 0$ and is model dependent. The temperature $T_D$ is then
\begin{equation}
T_D \sim \left[\frac{J_{\mbox{dc}}D}{\mathcal{N}^{5/3}}\right]^{1/(n+1)}.
\end{equation}
In the large-$\mathcal{N}$ limit the temperature in the nonequilibrium steady state approaches zero.

The presence of a nonzero $T_{D}$ results in an exponential decay for the current oscillations at large times, for in that case the ohmic kernel is $\Sigma^{K}(\tau)\sim \pi^2 T_{D}^{2}/\sinh^{2}(\pi T_{D}\tau).$ This goes like $1/\tau^{2}$ for $\tau \ll 1/T_{D}$ but decays exponentially as $T_{D}^{2}e^{-2\pi T_{D}\tau}$ for $\tau \gg 1/T_{D}$. 
The decay time for the oscillations is large due to large $\mathcal{N}$, which offers a rather large time window where the $1/\tau^{2}$ decay of the oscillations can be observed. However ultimately for $\tau \gg 1/T_{D},$ the current oscillations will decay exponentially. 
 
In the following Section, we study the effect of higher
order (in $g$) processes on the steady state part of the current. These higher order processes govern the dc current response 
at small values of $D.$  

\section{DC current at low fields: higher order processes\label{sec:Steady-state}}

Here we are interested in the long time steady state response here, for
which we turn on the electric field at $t=-\infty$ and for all later
times, the vector potential is simply $A_{k,1}(t)=Dt$ (i.e. without
the theta function in time). In this case, the expression for current
given in Eq. (\ref{eq:J-func}) assumes a simpler form, 
\begin{align}
J & =2\iota g\int d\tau\left[e^{-iD\tau}\Pi^{+-}(\tau)L^{+-}(\tau)-e^{iD\tau}\Pi^{-+}(\tau)L^{-+}(\tau)\right],\label{eq:J-steady}
\end{align}
since the terms involving Eq. (\ref{eq:J-func}) involving the bond
correlators $\Pi^{++}$ and $\Pi^{--}$ cancel out. Furthermore, for
a given sign of $D,$ only one of the two terms in the integrand contributes.
In the rest of the paper, we will assume $D>0$ unless otherwise stated,
and in this case, only the first term in the integrand in Eq. (\ref{eq:J-steady})
needs to be calculated. The perturbative expansion of $J$ is now
obtained by expanding the bond correlators in increasing orders in
$g,$ 
\begin{align*}
\Pi_{\sigma,\sigma'} & =\Pi_{\sigma,\sigma'}^{(0)}+\Pi_{\sigma,\sigma'}^{(1)}+\cdots.
\end{align*}
From Sec. \ref{sec:Transient-current-response}, we have the leading
order contribution to current as $J^{(1)}=(g/\pi)[(D-2E_{C})\Theta(D-2E_{C})-(D+2E_{C})\Theta(-D-2E_{C})].$
We now consider the contribution to the current in the second order
in the tunneling conductance $g.$ 

\subsection{Second order steady state response}

The first order correction to the bare bond correlator of the link
labeled $(k,1)$ is 
\begin{align}
\Pi_{\mu,\mu'}^{(1)}(\tau,\tau') & =\iota g\sum_{n,\sigma\sigma'}\int_{t,t'}W_{\mu\mu'\sigma\sigma'}^{k,n}(\tau,\tau',t,t')L^{\sigma\sigma'}(t-t')e^{-\iota D(t-t')},\label{eq:Pi-1}
\end{align}
where 
\begin{align}
W_{\mu\mu'\sigma\sigma'}^{k,n}(\tau,\tau',t,t') & =\left<\exp\left[-\iota\phi_{k,1}^{\mu}(\tau)+\iota\phi_{k,1}^{\mu^{'}}(\tau^{'})-\iota\phi_{n,1}^{\sigma}(t)+\iota\phi_{n,1}^{\sigma^{'}}(t^{'})\right]\right>_{S_{C}}.\label{eq:W-1}
\end{align}
Let us define the four-point site correlators, 
\begin{equation}
C_{\mu\mu^{'}\sigma\sigma^{'}}(\tau,\tau^{'},t,t^{'})=\left<\exp\left[-\iota\phi^{\mu}(\tau)+\iota\phi^{\mu^{'}}(\tau^{'})-\iota\phi^{\sigma}(t)+\iota\phi^{\sigma^{`}}(t^{'})\right]\right>_{S_{C}}.\label{eq:C-4pt}
\end{equation}
We now express the function $W_{\mu\mu'\sigma\sigma'}^{k,n}(\tau,\tau',t,t')$
in terms of the two and four point site correlators. For $n=k\pm1,$ 

\begin{equation}
W_{\mu\mu'\sigma\sigma'}^{k,n}(\tau,\tau',t,t')=C_{\sigma'\sigma}(t'-t)C_{\mu'\mu\sigma\sigma'}(\tau',\tau,t,t')C_{\mu\mu'}(\tau-\tau'),\label{W-2}
\end{equation}
while for $n=k,$

\begin{equation}
W_{\mu\mu'\sigma\sigma'}^{k,n}(\tau,\tau',t,t')=C_{\mu'\mu\sigma'\sigma}(\tau',\tau,t',t)C_{\mu\mu'\sigma\sigma'}(\tau,\tau',t,t').\label{eq:W-3}
\end{equation}
The correlator $W$ is nonzero only for $n=k\pm1\,k.$ For the calculation
of current we only need the $W_{\mu\mu'\sigma\sigma'}$ with $\mu,\mu'=\{+,-\}.$
These involve the following four-point site correlators: 
\begin{align}
C_{+-++}(\tau,\tau^{'},t,t^{'})= & \exp\left[-\iota E_{C}\left(-|t-\tau|+|t^{'}-\tau|+|t-t^{'}|-t-\tau+t^{'}+\tau^{'}\right)\right],\\
C_{+-+-}(\tau,\tau^{'},t,t^{'})= & \exp\left[-\iota E_{C}\left(-|t-\tau|+|t^{'}-\tau^{'}|-2(t-t^{'}+\tau-\tau^{'})\right)\right],\\
C_{+--+}(\tau,\tau^{'},t,t^{'})= & \exp\left[-\iota E_{C}\left(|t^{'}-\tau|-|t-\tau^{'}|\right)\right],\\
C_{+---}(\tau,\tau^{'},t,t^{'})= & \exp\left[-\iota E_{C}\left(-|t-\tau^{'}|+|t^{'}-\tau^{'}|-|t-t^{'}|-t-\tau+t^{'}+\tau^{'}\right)\right],\\
C_{-+++}(\tau,\tau^{'},t,t^{'})= & \exp\left[-\iota E_{C}\left(|t-\tau^{`}|-|t^{'}-\tau^{'}|+|t-t^{'}|+t+\tau-t^{'}-\tau^{'}\right)\right],\\
C_{-++-}(\tau,\tau^{'},t,t^{'})= & \exp\left[-\iota E_{C}\left(|t-\tau^{'}|-|t^{'}-\tau|\right)\right],\\
C_{-+-+}(\tau,\tau^{'},t,t^{'})= & \exp\left[-\iota E_{C}\left(|t-\tau|-|t^{'}-\tau^{'}|+2(t-t^{'}+\tau-\tau^{'})\right)\right],\\
C_{-+--}(\tau,\tau^{'},t,t^{'})= & \exp\left[-\iota E_{C}\left(|t-\tau|-|t^{'}-\tau|-|t-t^{'}|+t+\tau-t^{'}-\tau^{'}\right)\right].
\end{align}
The four-point site correlators clearly satisfy the identities 
\begin{align}
C_{\mu\mu^{'}\sigma\sigma^{'}}(\tau,\tau^{'},t,t^{'}) & =C_{\sigma\sigma'\mu\mu'}(t,t^{'},\tau,\tau^{'}),\nonumber \\
C_{\mu\mu^{'}\sigma\sigma^{'}}(\tau,\tau^{'},t,t^{'}) & =C_{\bar{\mu}\bar{\mu}'\bar{\sigma}\bar{\sigma}'}(\tau,\tau',t,t'),\label{eq:4pt-id}
\end{align}
where the bar on the subscripts interchanges the $+$ and $-$ indices.

From the structure of the four-point site correlators, we see that
the expression for the bond correlators has nonanalytic terms of the
type $e^{\iota E_{C}|t_{1}-t_{2}|}.$ To deal with these, we make
use of the identity,

\[
e^{-\iota E_{C}|t|}=\lim_{\eta\rightarrow0}\frac{\iota E_{C}}{\pi}\int_{-\infty}^{\infty}\frac{d\omega\,e^{-\iota\omega t}}{(\omega-E_{C}+\iota\eta)(\omega+E_{C}-\iota\eta)}.
\]
We then express $L^{\sigma\sigma'}(t-t')$ in the Fourier basis and
then perform the $t,\,t'$ integrals in Eq.(\ref{eq:Pi-1}). After
some effort we get the following expression for $\Pi_{+-}^{(1)}:$

\begin{align}
\Pi_{+-}^{(1)}(\tau) & =\frac{4\iota E_{C}^{2}g}{\pi}\lim_{\eta\rightarrow0}\int d\omega\left[\frac{L^{+-}(\omega-D)e^{\iota2E_{C}\tau}(e^{-\iota\omega\tau}-1)}{(\omega^{2}+\eta^{2})((\omega-2E_{C})^{2}+\eta^{2})}\right.\nonumber \\
 & \qquad\qquad\qquad\qquad\qquad+\frac{H^{+-}(\omega-D)e^{i2E_{C}\tau}(1-e^{\iota(4E_{C}+\omega)\tau})}{((\omega+4E_{C})^{2}+\eta^{2})((\omega+2E_{C})^{2}+\eta^{2})}\nonumber \\
 & \qquad\qquad\qquad\qquad\qquad+\frac{2L^{+-}(\omega-D)e^{\iota2E_{C}\tau}(e^{-\iota(\omega-6E_{C})\tau}-1)}{((\omega-6E_{C})^{2}+\eta^{2})((\omega-2E_{C})^{2}+\eta^{2})}\nonumber \\
 & \qquad\qquad\qquad\qquad\qquad\left.+\frac{2H^{+-}(\omega-D)(e^{\iota2E_{C}\tau}-e^{\iota\omega\tau})}{((\omega-2E_{C})^{2}+\eta^{2})((\omega+2E_{C})^{2}+\eta^{2})}\right],\label{eq:Pi1-final}
\end{align}
where $H^{+-}(\omega)=\Sigma^{+}(\omega)-\Sigma^{-}(\omega)+\Sigma^{K}(\omega).$
Using Eq. (\ref{eq:Pi1-final}) in Eq. (\ref{eq:J-steady}), we obtain
the second order contribution to the current:
\begin{align}
J^{(2)} & =-\frac{8g^{2}E_{C}^{2}}{\pi}\lim_{\eta\rightarrow0}\int d\omega\left[\frac{L^{+-}(\omega-D)[L^{+-}(2E_{C}-D-\omega)-L^{+-}(2E_{C}-D)]}{(\omega^{2}+\eta^{2})((\omega-2E_{C})^{2}+\eta^{2})}\right.\nonumber \\
 & \qquad\qquad\qquad\qquad\qquad+\frac{H^{+-}(\omega-D)[L^{+-}(2E_{C}-D)-L^{+-}(\omega+6E_{C}-D)]}{((\omega+4E_{C})^{2}+\eta^{2})((\omega+2E_{C})^{2}+\eta^{2})}\nonumber \\
 & \qquad\qquad\qquad\qquad\qquad+2\frac{L^{+-}(\omega-D)[L^{+-}(8E_{C}-D-\omega)-L^{+-}(2E_{C}-D)]}{((\omega-6E_{C})^{2}+\eta^{2})((\omega-2E_{C})^{2}+\eta^{2})}\nonumber \\
 & \qquad\qquad\qquad\qquad\qquad\left.+2\frac{H^{+-}(\omega-D)[L^{+-}(2E_{C}-D)-L^{+-}(\omega-D)]}{((\omega-2E_{C})^{2}+\eta^{2})((\omega+2E_{C})^{2}+\eta^{2})}\right].\label{eq:J2-1}
\end{align}
From the step-like structure of the $L^{+-}$ and $H^{+-}$ functions,
we find that $J^{(2)}=0$ for $D<E_{C};$ thus, $J^{(2)}$ has a smaller
threshold compared to $J^{(1)},$ which vanishes below $2E_{C}.$
For $E_{C}\leq D<2E_{C},$ the calculation of the current simplifies
considerably since only one term makes a nonzero contribution in Eq.
(\ref{eq:J2-1}), and we have 
\begin{align}
J^{(2)} & =-\frac{8g^{2}E_{C}^{2}}{\pi}\lim_{\eta\rightarrow0}\int d\omega\frac{L^{+-}(\omega-D)L^{+-}(2E_{C}-D-\omega)}{(\omega^{2}+\eta^{2})((\omega-2E_{C})^{2}+\eta^{2})},\quad E_{C}\leq D<2E_{C},\label{eq:J2-2}
\end{align}
and upon performing the integration we arrive at 
\begin{align}
J^{(2)} & =\frac{2g^{2}}{\pi^{3}E_{C}}\left((D-E_{C})^{2}+E_{C}^{2}\right)\log\left[\frac{D^{2}}{(D-2E_{C})^{2}}\right]-\frac{8g^{2}}{\pi^{3}}(D-E_{C}),\quad E_{C}\leq D<2E_{C}.\label{eq:J2-3}
\end{align}
Just above the threshold for $J^{(2)}$, $D=E_{C},$ the current has
a power-law behavior, 
\begin{align}
J^{(2)} & \approx\frac{8g^{2}E_{C}}{\pi^{3}}\left(\frac{D}{E_{C}}-1\right)^{3},\label{eq:J2-thresh}
\end{align}
which is to be contrasted with the linear behavior of $J^{(1)}$ above
its threshold. At the other end, $D=2E_{C},$ the expression for $J^{(2)}$
has a logarithmic divergence. Physically, this is a manifestation
of a resonance: $D=2E_{C}$ is the condition for creating a particle-hole
dipole excitation in neighboring grains. For higher fields, $D>2E_{C},$
more terms in Eq. (\ref{eq:J2-1}) will now contribute to $J^{(2)};$
however, none of these terms eliminate the logarithmic singularity. 

The second order perturbation correction to the current is justified
provided one does not get too close to the singular point, i.e., 
\begin{align}
g\ln\left|\frac{D}{2E_{C}-D}\right| & \lesssim1.\label{eq:pert-th-criterion}
\end{align}
Similar logarithmic divergence is also evident in $\Pi^{(1)}(\tau).$
On the other hand, the bond correlator, $\Pi=\Pi^{(0)}+\Pi^{(1)}+\cdots,$
by definition is bounded by $\pm1.$ This clearly shows that the divergence
in current at the resonance is the result of a perturbative treatment
about the bare charging action. The region of validity of the perturbative
treatment could be increased in principle by a resummation of the
leading singular terms to all orders in $g.$ Unfortunately, the number
of processes contributing to current in higher orders increases rapidly
with the order, rendering the calculation of the current at intermediate
fields (sufficiently larger than the lowest threshold) quite complicated.
The other possibility is a phase transition from the Mott phase to
a conducting, metallic phase whose boundary is given by the condition
$g\ln(2E_{C}/\epsilon)=1,$ with $\epsilon=2E_{C}-D\ll E_{C}.$ The
resummation and possible phase transition will be studied in detail
elsewhere. Incidentally, the energy scale $\epsilon=E_{C}e^{-1/g}$
also appears in the scaling analysis of the single site equilibrium
AES model close to the degeneracy point, $n_{g}=1/2$ \cite{falci1995unified}.
Below this scale, phase fluctuations renormalize the gate charge to
the fixed point value, $n_{g}=1/2,$ which corresponds to resonant
transmission. Finally, for very small values of $2E_{C}-D,$ we expect
that the energy level discreteness of the dots will begin to matter,
and at resonance, the lower cutoff for $|2E_{C}-D|$ should at least
be of the order of the mean level spacing $\delta,$ i.e., we need
$g<1/\ln(2E_{C}/\delta).$ 

\subsection{Higher order contributions and current response at low fields}

At low fields, finite contributions to the current appear only at
higher orders. An order-$n$ process has a threshold field $D_{\text{th}}^{(n)}=2E_{C}/n.$
Physically, a large-distance cotunneling process provides the potential
energy gain required to overcome Coulomb blockade. During the cotunneling
process between sites labeled $i$ and $i+n,$ the classical charges,
$n^{c},$ at the $n-1$ intermediate sites only have virtual transitions
and thus the only Coulomb blockade cost appears at the sites $i$
and $i+n.$ The pure cotunneling process gives the lowest threshold
value, $D_{\text{th}}^{(n)}$, at any order. The contribution to the
current from this process can be shown to be 
\begin{align}
J^{(n)} & =\iota2ng\left(\frac{\iota2gE_{C}^{2}}{\pi}\right)^{n-1}K^{(n)},\label{eq:Jn-1}
\end{align}
where
\begin{align}
K^{(n)} & =\int\prod_{i=1}^{n-1}d\omega_{i}\left[\prod_{j=1}^{n-1}\frac{L^{+-}(\omega_{j}-D)}{\omega_{j}^{2}(\omega_{j}-2E_{C})^{2}}\right]L^{+-}(2E_{C}-D-\sum_{p=1}^{n-1}\omega_{p}).\label{eq:Kn}
\end{align}
The $L^{+-}$ functions constrain the frequency integration and we
have 
\begin{align}
K^{(n)} & =\left(\frac{\iota}{2\pi}\right)^{n}\int_{2E_{C}-(n-1)D}^{D}d\omega_{1}\int_{2E_{C}-(n-2)D-\omega_{1}}^{D}d\omega_{2}\,\,\,\cdots\int_{2E_{C}-D-\sum_{p=1}^{n-2}\omega_{p}}^{D}d\omega_{n-1}\nonumber \\
 & \times\frac{(\omega_{1}-D)(\omega_{2}-D)\cdots(\omega_{n-1}-D)(2E_{C}-D-\sum_{p=1}^{n-1}\omega_{p})}{\omega_{1}^{2}\ldots\omega_{n-1}^{2}(\omega_{1}-2E_{C})^{2}\ldots(\omega_{n-1}-2E_{C})^{2}}.\label{eq:Kn-2}
\end{align}
The integral gets the dominant contribution from the vicinity of $\omega_{i}=D,$
and is approximately 
\begin{align}
K^{(n)} & \approx\left(-\frac{\iota}{2\pi}\right)^{n}\frac{n^{(2n-1)}}{(2n-1)!}\frac{(D-D_{\text{th}}^{(n)})^{(2n-1)}}{D^{2(n-1)}(2E_{C}-D)^{2(n-1)}}\Theta(D-D_{\text{th}}^{(n)}),\quad\frac{D-D_{\text{th}}^{(n)}}{D_{\text{th}}^{(n)}}\ll1.\label{eq:Kn-3}
\end{align}
Combining Eqs. (\ref{eq:Jn-1}) and (\ref{eq:Kn-3}), and making the
Stirling approximation for factorials, we obtain, for large $n,$
\begin{align}
J^{(n)} & \sim ng^{n}\left(\frac{e}{2\pi}\right)^{2n-1}\left(\frac{2E_{C}}{D(2E_{C}-D)}\right)^{2(n-1)}\left(D-D_{\text{th}}^{(n)}\right)^{2n-1}\Theta(D-D_{\text{th}}^{(n)})\label{eq:Jn-final}\\
 & \approx anb^{n}D\left(1-\frac{n_{D}}{n}\right)^{2n-1}\Theta\left(1-\frac{n_{D}}{n}\right),\label{eq:Jn-final2}
\end{align}
where 
\begin{align}
a & =\frac{2\pi}{e}\left(1-\frac{1}{n_{D}}\right)^{2},\nonumber \\
b & =g\left(\frac{e}{2\pi}\right)^{2}\left(\frac{1}{1-n_{D}^{-1}}\right)^{2},\nonumber \\
n_{D} & =\frac{2E_{C}}{D}.\label{eq:Jn-parameters}
\end{align}
Denoting $[n_{D}]$ to be the least integer $\ge n_{D}$, the expression
for the total current is given by, 
\begin{equation}
J=\sum_{n=[n_{D}]}^{\infty}J^{(n)}.\label{jtot}
\end{equation}
For $D\ll E_{C}$, from the large $n$ form of $J^{(n)}$ in Eq. (\ref{eq:Jn-final2}),
we see that the expression for the total current is divergent for
$b\ge1$. We identify the onset of this divergence as the breakdown
of our perturbation theory which is developed to work in the Mott
phase and thus signals the nonequilibrium phase transition to a metallic
phase. Thus, for small values of the electric field, the phase boundary
for the nonequilibrium phase transition to this metallic phase is
given by setting $b=1:$ 
\begin{align}
g & =g_{0}\left[1-\frac{D}{2E_{C}}\right]^{2},\quad D\ll2E_{C},\label{eq:phase-bdry}
\end{align}
with $g_{0}$ a constant of order one. For given $g$ and $E_{C},$
the critical electric field is 
\begin{align}
D_{c} & =2E_{C}(1-\sqrt{g/g_{0}}).\label{eq:crit-field}
\end{align}

Let us now look into the form of current within the Mott phase for
small $D$. From Eq.(\ref{eq:Jn-final2}), we see that the expression
for $J$ in eq.(\ref{jtot}) can be approximated by a saddle point
approximation if $b\ll1$. For this we first rewrite Eq. (\ref{jtot})
as 
\begin{equation}
J=aD\sum_{n=[n_{D}]}^{\infty}\exp\left[\ln n+n\ln b+(2n-1)\ln\left(1-\frac{n_{D}}{n}\right)\right].
\end{equation}
The saddle point condition is (neglecting some small terms): 
\begin{equation}
\ln b+2\ln\left(1-\frac{n_{D}}{n}\right)+\frac{2n_{D}}{n-n_{D}}=0.
\end{equation}
In terms of $x=n_{D}/n$, an approximate solution of the above equation
can be written as 
\begin{equation}
x=x^{*}-(1-x^{*})^{2}\ln(1-x^{*}),
\end{equation}
where, $x^{*}=\left(1-\frac{2}{\ln b}\right)^{-1}$. The form of current
then turns out to be (for $D<D_{c}$), 
\begin{align}
J & \sim a\begin{cases}
D\exp\left[-(4E_{C}/D)\ln\left[\sqrt{\frac{g_{0}}{g}}\left(1-\frac{D}{2E_{C}}\right)\right]\right], & D\ll D_{c}\\
D_{c}\left(\frac{D_{c}}{D_{c}-D}\right)^{2}, & \frac{D_{c}-D}{D_{c}}\ll1.
\end{cases}\label{eq:J-small-D}
\end{align}
Thus as the critical field $D_{c}$ is approached, the perturbation
series for the current diverges, signaling the breakdown of the Mott
insulator state. Farther away from the critical field, the form of the current
resembles an activated behavior, with the driving field $D$ taking the role
of the temperature, and the Arrhenius cost changed from the bare value $E_C$ to 
an effectively (field-dependent) lower value, $E^{\text{eff}}_C = E_C \ln\left[\sqrt{\frac{g_{0}}{g}}\left(1-\frac{D}{2E_{C}}\right)\right]^4.$ The similarity with thermal activation is not surprising since the constant and uniform electric field 
also generates free particles across the excitation gap, albeit through the Landau-Zener-Schwinger mechanism. 
Closer to the transition field  $D_c,$ we expect the divergence of the current response in Eq. (\ref{eq:J-small-D}) to ultimately
get cut off by processes we did not take into account in our perturbation series that consisted, at every order, of only the respective threshold contributions. Further work needs to be done to establish if there is any non-analyticity in the current response across the transition field, for that would imply a true nonequilibrium phase transition and not a crossover between the Mott insulator and bad metal phases.

\section{Discussion \label{sec:discussion}}
In summary, we developed an effective Keldysh field theory for studying the nonequilibrium response of dissipative Mott insulator systems, and used it to study the nonequilibrium
current response to a uniform electric field switched on at some instant of time. Our model, a Keldysh generalization of the AES model for Mott insulators, is in effect a 
bosonization of the Hubbard model with a large number ($\mathcal N$) of electron flavors at the lattice sites. The effective degrees of freedom are the excess charges at the 
sites and the phases conjugate to these. The large-$\mathcal N$ is
simultaneously a source of dissipation through the Landau damping mechanism and also affords significant simplification of the effective action (in comparison with the usual
Hubbard model) by suppressing all terms that are higher than second order in the interdot tunneling amplitude. 

The quantum effect that survives in the large-$\mathcal N$ limit is charge quantization, which is respected at every stage in the analysis of our problem. 
The charge quantization is reflected in sustained Bloch-like oscillations that decay as an inverse square power-law in time up to a large time scale $\tau_{D} \sim 1/T_{D} \sim \mathcal{N}^{-\alpha},\,\alpha>0.$ 
The effect of correlations is to split the Bloch oscillation frequency into two beating frequencies whose difference is of the order of the Coulomb repulsion scale.

The power-law decay of the current oscillations signifies the persistence of Coulomb blockade or charge quantization effects. At small values of tunneling $g,$ Coulomb blockade effects dominate and the dot charge fluctuations are weak. The presence of a large number of energy levels in the dots may scramble the phase of the electronic states but is not able to erase charge quantization effects at weak tunneling.  
In the effective action, $\Sigma^{K}(t,t')$ that represents correlations between tunneling events at time $t$ and $t',$ decays
as a power-law in the zero temperature limit, $1/(t-t')^2,$ which is essentially why our current oscillations at weak tunneling obey the same power-law decay. But the energy dissipation that is responsible for a dc component in the current, also results in a finite system temperature $(T_{D})$, which is determined by a combination of the power dissipation and the (weak) coupling to the external heat bath. Note that $T_{D}$ is very small due to the effect of large-$\mathcal{N}$. The non-zero $T_{D}$ causes the $1/t^{2}$ decay to crossover to an exponential decay after $t\sim 1/T_{D}$.

When $g$ is large, the behavior is very much like a classical RC circuit where charge can take continuous values. In this regime, the charge fluctuations are strong (i.e., charge quantization effects are weak), and 
the current oscillations decay exponentially. To see this, at large $g$, we expand 
around the saddle point configuration of the effective action (Eq. (\ref{eq:stun-pm})) and to quadratic order in the phase fields.
In the classical-quantum space, the action takes the form,
\begin{equation}
  S[\phi]=\sum_{q}\int_{\omega} \begin{bmatrix}\phi^{c} & \phi^{q}
 \end{bmatrix}_{(q,\omega)} \begin{bmatrix} 0 & -\frac{\omega^{2}}{E_{C}}-i\omega h(q) \\
 -\frac{\omega^{2}}{E_{C}}+i\omega h(q) & 2ih(q)|\omega|
 \end{bmatrix} \begin{bmatrix}\phi^{c}\\
\phi^{q} 
\end{bmatrix}_{(-q,-\omega)},
\end{equation}
where $h(q)=2g(1-\cos q)$.
The inverse of the matrix is given by
\begin{equation}
 \frac{-1}{(h(q))^{2}\omega^{2}+\frac{\omega^{4}}{E_{C}^{2}}}\begin{bmatrix} 2ih(q)|\omega|  & \frac{\omega^{2}}{E_{C}}+i\omega h(q) \\
 \frac{\omega^{2}}{E_{C}}-i\omega h(q) & 0\end{bmatrix}
\end{equation}
We thus have
\begin{equation}
 G^{\pm}(\omega)=-\frac{1}{\omega\left( \frac{\omega}{E_{C}}\mp ih(q)\right)}, \,\,
 G^{K}(\omega)=-\frac{2ih(q)|\omega|}{(h(q))^{2}\omega^{2}+\frac{\omega^{4}}{E_{C}^{2}}}.
\end{equation}
The presence of the imaginary pole in the retarded Green's function implies in the time domain it has an exponential decay with a characteristic time scale, $1/(gE_{C}),$ which is equivalent to the time constant, $\tau=RC,$ of a resistively shunted capacitor.
So the oscillations in the current after the source is turned on at $t=0$ would also decay exponentially with the same characteristic time scale. 

A major challenge in the area has been to demonstrate a DC current response in lattice translationally invariant Hubbard models. We identified the role played by
dissipation in suppressing the Bloch oscillations (even if as a power law in time) and enabling a finite DC current response. We analyzed the DC current response
taking into account higher order cotunneling processes that allow a trade-off between the reduced probability of a long-distance cotunneling and energy gain from
the applied electric field. The response at small electric fields is found to be of the LZS form, $J \sim D [g/\ln^2(1/g)]^{2E_C/D},$ although the exponent is
proportional to the Mott gap $E_C$ instead of the usual $e^{-E_{C}^{2}/D}$ expected for pair-production probability in the dissipation-free case \cite{oka2005}. 
We do not find a threshold field below which DC conduction is absent since at any small field, DC conduction is possible through sufficiently high order cotunneling.
At higher fields, the perturbation expansion of the current in powers of the small tunneling conductance breaks down, and from this we obtain the phase boundary for 
the electric field driven Mott insulator to a conducting state. Both a phase tranisiton and a (rapid) crossover are consistent with our results, since the expression that we have obtained for higher order contributions to the current is only valid at very small values of the driving field $D;$ but the instablity at $D_c = 2 E_C (1-\sqrt{g/g_0})$ in Eq. (105) suggested by the 
divergence of perturbation theory occurs at a value of $D$ that is not necessarily small. Thus corrections to our expressions might become relevant 
in actually determining  whether there is a real phase transition or not.

The AES model regards the interdot tunneling processes to be of the Fermi Golden-Rule type, which breaks down when the characteristic energies of particle-hole
excitations in the dots approach the mean level spacing, $\delta.$ Therefore the typical potential drop between neighboring sites or the temperature should 
exceed $\delta.$ This imposes a cutoff on the regime of validity of our analysis.

We conclude with a brief discussion of future directions. Our approach can also be useful for the study of other far from equilibrium problems of current interest. 
For example, it is an interesting question as to how 
an initial non-thermal distribution of dot charges would evolve with time - in particular whether the long-time behavior retains any memory of the initial conditions.
Similar questions have been posed, for example, in the context of relaxation of initial charge disctribution in bosonic cold atom systems \cite{bloch2017probing} and 
the approach to thermal equilibrium in fermionic quantum chains \cite{bulchandani2017solvable}. Our Keldysh-AES model can also be used to study
the energy transport. The problem we have attacked in our paper is the current response to a uniform DC electric field; however,
the approach is readily generalized to problems involving time-dependent drives. In this context, it would be interesting to compare with periodically driven
Hubbard chains in the absence of dissipation \cite{tsuji2012repulsion}. 
As we noted in our paper, there are two special values of the background charge on a dot - integer and half odd integer. 
The integer case that we studied in detail
corresponds to a Mott insulator, while the latter is a correlated ``bad'' metal. The nonequilibrium response close
to half odd integer background charges is an open question. 
Another interesting direction would be to study the nonequilibrium response of driven Josephson-junction arrays.
This direction, especially after taking into account long-range Coulomb interactions, would shed more light to understand the sudden
jumbs observed in the I-V characteristics of disordered superconductors that are in the insulating side and in the proximity of 
superconductor to insulator transition \cite{vinokur2008superinsulator,ovadia2015evidence,mironov2018charge,sankar2018disordered}.

\section{Acknowledgements}

The authors are delighted to thank Rajdeep Sensarma and Gautam Mandal for illuminating discussions. V.T. would like to 
thank the Department of Science and Technology, Govt. of India,
for a Swarnajayanti Grant (Grant No. DST/SJF/PSA-0212012-13).

\appendix

\section{Normalization of the partition function}

A key property of the Keldysh partition function is that in the absence
of source fields, the partition function is normalized. Demonstrating
this for the Keldysh-AES action requires one to take into account
the correct causal structure of the Green functions. We expand $\exp[\iota S_{\mbox{tun}}[\phi]]$
in powers of $g$. To leading order, we get,

\begin{equation}
Z^{(0)}=\int[D\phi][Dn]\exp\left[\iota\left(S_{C}[n,\phi]\right)\right]
\end{equation}
Doing the functional integration over $\phi$, we see that the constraints
$\partial_{t}n^{+}=0$ and $\partial_{t}n^{-}=0$ are imposed and
then it immediately follows from the boundary condition , $n^{+}(-\infty)=n^{-}(-\infty)$,
that $Z^{(0)}=1$. 

\subsubsection*{Order g }

\begin{equation}
Z^{(1)}=\iota g\sum_{k}\int_{-\infty}^{\infty}\int_{-\infty}^{\infty}\mbox{ dt }\mbox{ dt' }L^{\sigma\sigma^{'}}(t-t')\left<\exp\left[-\iota\phi_{k,1}^{\sigma}(t)+\iota\phi_{k,1}^{\sigma'}(t')\right]\right>_{0},
\end{equation}
where $<>_{0}$ denotes averaging with respect to the bare action.
Thus, 
\begin{equation}
Z^{(1)}=\iota g\sum_{k}\int_{-\infty}^{\infty}\int_{-\infty}^{\infty}\mbox{ dt }\mbox{ dt' }L^{\sigma\sigma^{'}}(t-t')\Pi_{\sigma\sigma^{'}}(t-t^{'}).
\end{equation}
We have the site correlators, 
\begin{eqnarray}
C_{++}(t-t^{'}) & = & \exp\left[-\iota E_{C}|t-t^{'}|\right],\\
C_{+-}(t-t^{'}) & = & \exp\left[\iota E_{C}(t-t^{'})\right],\\
C_{-+}(t-t^{'}) & = & \exp\left[-\iota E_{C}(t-t^{'})\right],\\
C_{--}(t-t^{'}) & = & \exp\left[\iota E_{C}|t-t^{'}|\right].
\end{eqnarray}
Then we get the bond correlators, 
\begin{eqnarray}
\Pi_{++}(t-t^{'}) & = & \exp\left[-2\iota E_{C}|t-t^{'}|\right],\\
\Pi_{+-}(t-t^{'}) & = & \exp\left[2\iota E_{C}(t-t^{'})\right],\\
\Pi_{-+}(t-t^{'}) & = & \exp\left[-2\iota E_{C}(t-t^{'})\right],\\
\Pi_{--}(t-t^{'}) & = & \exp\left[2\iota E_{C}|t-t^{'}|\right].
\end{eqnarray}
From the bond correlators we immediately see that the term involving
$\Sigma^{K}$ vanishes. Now lets look at the term with $\Sigma^{+}$.
In the time representation, we have to keep in mind that it comes
with the causality factor $\theta(t)$ and hence we write it as $\Sigma(t)\theta(t)$.
The term involving this reads as, 
\begin{equation}
\Sigma^{+}(t)\theta(t)\left[\exp(\iota E_{C}t)-\exp(-\iota E_{C}t)-\exp(\iota E_{C}|t|)+\exp(-\iota E_{C}|t|)\right].
\end{equation}
Because of the presence of the Theta function, we see that we can
remove the modulus sign from the last two terms and then clearly this
contribution vanishes. Similarly we see that the contribution from
terms involving $\Sigma^{-}$ also vanishes. Hence we see that the
order $g$ contribution to the partition function vanishes. We assume
that all the higher order $g$ contributions to the partition function
vanishes too and thus the partition function is truly equal to 1.



\end{document}